\documentclass[%
 aip, amsmath,amssymb, reprint, floatfix
]{revtex4-1}

\usepackage{graphicx}
\usepackage{dcolumn}
\usepackage{bm}

\usepackage[utf8]{inputenc}
\usepackage[T1]{fontenc}
\usepackage{mathptmx}
\usepackage{etoolbox}
\usepackage{soul}

\makeatletter
\def\@email#1#2{%
 \endgroup
 \patchcmd{\titleblock@produce}
  {\frontmatter@RRAPformat}
  {\frontmatter@RRAPformat{\produce@RRAP{*#1\href{mailto:#2}{#2}}}\frontmatter@RRAPformat}
  {}{}
}%
\makeatother

\usepackage{empheq}
\draft 

\usepackage{bm}
\usepackage{xcolor}
\usepackage{tikz}
\usetikzlibrary{positioning}
\usetikzlibrary {arrows.meta}
\newcommand{\curlyL}{\mathcal{L}}

\newcommand{\INTL}{\int\limits_{-L}^L}

\usepackage{chemformula}

\begin{document}

\title{Turning catalytically active pores into active pumps}

\author{G. C. Antunes}
\affiliation{Helmholtz-Institut Erlangen-Nürnberg für Erneuerbare Energien (IEK--11), Forschungszentrum J\"ulich, Cauer Str.~1, 91058 Erlangen,\,Germany}

\author{P. Malgaretti}
 \affiliation{Helmholtz-Institut Erlangen-Nürnberg für Erneuerbare Energien (IEK--11), Forschungszentrum J\"ulich, Cauer Str.~1, 91058 Erlangen,\,Germany}

 \author{J. Harting}
\affiliation{Helmholtz-Institut Erlangen-Nürnberg für Erneuerbare Energien (IEK--11), Forschungszentrum J\"ulich, Cauer Str.~1, 91058 Erlangen,\,Germany}
\affiliation{Department Chemie- und Bioingenieurwesen und Department Physik, Friedrich-Alexander-Universit\"at Erlangen-N\"urnberg, F\"{u}rther Stra{\ss}e 248, 90429 N\"{u}rnberg, Germany}

\date{\today}

\begin{abstract}
We develop a semi-analytical model of self-diffusioosmotic transport in active pores, which includes advective transport and the inverse chemical reaction which consumes solute. In previous work (Phys. Rev. Lett. 129, 188003, 2022), we have demonstrated the existence of a spontaneous symmetry breaking in fore-aft symmetric pores that enables them to function as a micropump. We now show that this pumping transition is controlled by three timescales. Two timescales characterize advective and diffusive transport. The third timescale corresponds to how long a solute molecule resides in the pore before being consumed. Introducing asymmetry to the pore (either via the shape or the catalytic coating) reveals a second type of advection-enabled transition. In asymmetric pores, the flow rate exhibits discontinuous jumps and hysteresis loops upon tuning the parameters that control the asymmetry. This work demonstrates the interconnected roles of shape and catalytic patterning in the dynamics of active pores, and shows how to design a pump for optimum performance. 
\end{abstract}

\maketitle 

\section{Introduction}

Over the last decades, novel methods to manipulate fluid flows on the micro- and nano-scale have enabled a technological leap \cite{Zhang2020,Hou2017,Huang2020,Squires2005,Whitesides2006}, with microfluidic devices being employed in fields such as chemical synthesis \cite{Bailey2021}, and biomedical engineering \cite{Yang2020}. The current need for increased sustainability has spawned a new field of chemistry concerned with reducing the environmental footprint of the chemical industry \cite{Zimmerman2020,Clark2012}. One way is to reduce the size of the chemical reactors themselves. It has been shown that chemical reactors on the scale of dozens to hundreds of micrometers can produce compounds in a way that is safer, faster, and more efficient when compared to traditional batch reactors\cite{Haswell2003, Kolb2004, DeWitt1999}. Microfluidic devices can also be used to produce and assemble nanoparticles\cite{Amreen2021,XDGH15}, or for 3D fabrication\cite{Warsi2018} via e.g. inkjet printing \cite{Guo2017}. Biomedical researchers often use microfluidic devices to e.g. develop novel drug delivery systems \cite{Egrov2021}, or to mimic biological systems such as the vascular system \cite{Ma2021,SFFVHM18}. Indeed, microfluidic chips may provide very quick and reliable diagnostics while requiring a minimal amount of biological sample from a patient \cite{Francesko2019}. 

A common need in all of these applications is the need to pump fluid in a controlled fashion. Thus, micropumps are commonly found in many microfluidic devices \cite{Wang2018}. A common challenge is the decreased effectiveness of traditional hydraulic pumping arising from the increased surface-to-volume ratio \cite{Laser2004}. An alternative is to use surface-driven flows, such as electroosmosis. This phenomenon occurs when a fluid in contact with a charged surface is pumped via an application of an external electric field \cite{Anderson1989}. The elevated voltage needed to operate an electroosmotic pump can however be an issue \cite{Hossan2018}. An alternative form of pumping that bypasses the need for an external field entirely is diffusioosmosis \cite{Shim2022}. When a solution is placed in contact with a solid phase such as the walls of a channel, solute-solid and solvent-solid interactions play a role. As the solvent is chemically distinct from the solute, these interactions will be of different strength and range. Thus, if the concentration of solute is spatially-varying, the overall interaction between solute and wall will also be spatially-varying, leading to local pressure imbalances near the solid. The end result is a fluid flow parallel to the solute concentration gradient \cite{Anderson1989}. One way to enforce an inhomogenous concentration of solute is by coating the solid with a catalyst, and to choose a chemically reactive solution. The solution is decomposed in the presence of the catalyst, generating a solute. Covering only part of the solid in catalyst ensures that the solute concentration is inhomogeneous. This technique has been used to fabricate self-motile Janus particles, which are half-coated in catalyst \cite{Ebbens_Review, Juliane_review, Howse2007, Popescu2016, Malgaretti2021}.  It has also been used to fabricate active, self-diffusioosmotic channels with self-pumping walls \cite{Yu2020}. Such diffusioosmotic channels can act as micropumps (i.e. exhibit a non-zero net flow rate) by breaking fore-aft symmetry either via their catalytic coating \cite{Michelin2019}, or by the shape of their corrugation \cite{Michelin2015}. Recently, it has been shown that breaking the channel's fore-aft symmetry by design is not necessary, as a spontaneous symmetry breaking occurs when the advection of solute plays a large enough role \cite{Antunes2022}. It is worth noting that the advection-enabled spontaneous symmetry breaking occurs for colloids as well, leading to diffusiophoretic isotropic colloids \cite{Michelin2013, deBuyl2013, Michelin2014}. The interplay of geometrical and chemical inhomogeneities are crucial for the performance of these active pumps, as it has been shown that, while flat channels with homogeneous chemical properties may exhibit local convection currents, they do not amount to a net flow \cite{Antunes2022, Michelin2020, Chen2021}.

Advection-enabled pumping in fore-aft symmetric channels is expected to occur when the timescale for advective transport matches the one of diffusive transport \cite{Antunes2022, Michelin2013}. Using typical experimental values for platinum-coated channels holding a hydrogen peroxide solution, this matching is expected to occur for channels on a scale of a hundred to a thousand micrometers \cite{Antunes2022}. To our knowledge however, there is no work regarding the effect of advection for a general channel (symmetric or otherwise), despite many microfluidic devices operating on precisely that scale. An understanding of transport in catalytically active materials is itself a subject of interest, as these materials play a large role in the chemical industry \cite{book_IndustrialCatalysis,Tanimu2022,Zhu2020, Pei2018, book_FuelCells}, and are expected to be vital for the decarbonization of society \cite{Anastas2002,Mayrhofer2014}. Transport on the scales ranging from micrometers to hundreds of micrometers is of great importance to the functioning of catalysts \cite{Stary2006,Selvam2014, He2012}, and it is at these scales where diffusioosmosis plays a role \cite{Howse2007,Yu2020}.

One often desires micropumps with flow rates that are stable over time. Thus, the most relevant properties of the active channel are those of its steady states. However, a confined system such as the active channel cannot reach a useful steady state unless some mechanism stops the product concentration from rising. Thus, implementations of the active pore as a micropump must include a sink of solute, regardless of its physical nature. This sink is often imposed in theoretical work by imposing a constant solute concentration in sections of the channel wall. This is unlikely to be the case in experimental setups. One way to ensure the existence of a steady state is to take into account the often-neglected inverse reaction that consumes solute. This reaction is always present \cite{Petlicki1998}, even if the rate may be significantly smaller than the rate of the forward reaction. Including microscopic reversibility has already been shown to alter the dynamics of active colloids \cite{Ryabov2022}. 

With this paper, we seek to shed light on the dynamics of active channels/pores by deriving a model of diffusioosmotic transport for corrugated pores. The chemical activity which drives the flows is spatially-varying, and the inverse chemical reaction is taken into account. Both advective and diffusive transport play a role, and our model captures the spontaneous symmetry breaking in fore-aft symmetric pores. We show that when the inverse chemical reaction is taken into account, the control parameter that controls the transition from non-pumping to pumping in symmetric channels is a combination of three timescales: the advective timescale, the diffusive timescale, and the time a solute molecule resides in the pore before being consumed. We further find a second type of advection-enabled transition. This transition occurs only for fore-aft asymmetric channels, which may exhibit bistability, a discontinuous jump in the flow rate, and hysteresis loops.

The remainder of this paper is structured as follows. Our model of the catalytically active pore is presented in Section~\ref{sec:model}. In Section~\ref{sec:FJ}, the governing equations are systematically reduced to an effective 1D model by employing the Fick-Jacobs and the lubrication approximations. The resulting model is employed in the study of a sinusoidally-shaped pore in Section~\ref{sec:sinusoidal}, which is split between fore-aft symmetric pores (Section~\ref{sec:sinusoidalSym}) and asymmetric pores (Section~\ref{sec:sinusoidalAsym}). We relax the condition of sinusoidal shape and examine pores with a ratcheted shape in Section~\ref{sec:NonSinusoidal}. Finally, in Section~\ref{sec:conclusions}, a summary of the results is given, and a connection to the experimental setup of platinum-coated pores containing a hydrogen peroxide solution is made.

\section{The model}

\label{sec:model}

\begin{figure*}[t]
	\centering
   \includegraphics[width=1.0\textwidth]{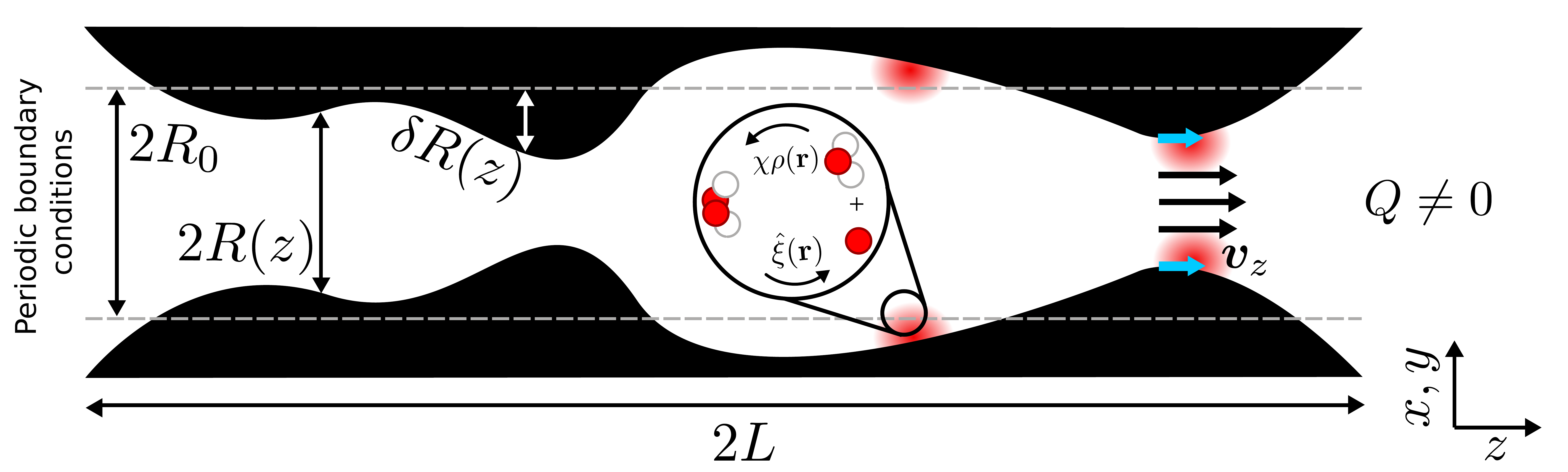}
   \caption{Cartoon of the longitudinal cross-section of the three-dimensional, axially-symmetric, active pore. The pore length is $2L$ and its variable radius is $R(z)$, with an average radius of $R_0$. The deviation from this average is $\delta R(z)$. The catalysis of the chemical solute occurs near the pore wall with a possibly spatially inhomogeneous rate $\hat{\xi}(\textbf{r})$. The deeper the color red, the higher the value of $\hat{\xi}(\textbf{r})$. The reverse chemical reaction removes solute in the bulk with a spatially homogeneous rate $\chi$ (see inset). The pore walls interact with the solution via the effective potential $U_{wall}$. A flow field $\bm{v}$ emerges due to an effective diffusioosmotic slip velocity (blue arrows), which may lead to the onset of a net non-zero flow rate $Q$.
   }
    	\label{fig:modelNotes}
\end{figure*}

 We consider a three-dimensional, axially symmetric pore with its axis along the z direction. The cross-section is thus circular with a radius which varies with z. The ends of the pore are located at $z = \pm L$. The pore walls are located at $x^2 + y^2=R^2(z)$, with the average radius $R_0$ defined as 
\begin{equation}
    R_0 = \frac{1}{2L}\int\limits_{-L}^L R(z) dz.
\end{equation}
Throughout this work, we will consider $R(z)$ to be a $2L$-periodic function in the $z$ direction. See Fig.~\ref{fig:modelNotes} for a sketch of the pore geometry. Inside this pore is a fluid that undergoes a chemical reaction due to the presence of catalytic material, which coats the pore walls inhomogeneously. We model this fluid as a mixture of three chemical species, the solvent, the reactant, and the product. In the vicinity of catalytic material, reactant is turned into product. The rate of production depends on factors such as the concentration of the catalytic material, and (in principle) of the reactant. However, in order to keep the model as simple as possible, in the following we specialize to the reaction-limited case. Indeed, this is the case for typical experimental setups \cite{Brown2014}. Accordingly, we assume the reactant concentration to be homogeneous in space. The concentration of solvent is then determined by the concentration of product, so as to keep the overall mixture incompressible.  In order to attain a steady state, the reactant must be replenished, and the product must be disposed of. In cases where the fluid is unbound, these processes occur due to diffusion from (for reactant) and into (for product) a chemical reservoir that is placed far away from the catalytic material \cite{Ibrahim2017}. 
In the case where the reaction takes place in a confined space, transfer of reactant (product) into (out of) the reactor has been obtained by exploiting the porosity of the confining walls~\cite{Palacci2010}. Regardless of its physical nature, a mechanism that replenishes reactant and disposes of product is needed to obtain a steady state, and thus steady pumping. We model such a mechanism by accounting for an inverse chemical reaction that converts the product back into reactant. This inverse reaction occurs with a rate $\chi \rho(\mathbf{r},t)$, where $\rho$ is the number density of the product, and $\chi$ is a sink constant of dimension $s^{-1}$. We further assume that the concentration of reactant is much larger than that of product, and thus does not appreciably change in time. As a consequence, the reaction that generates product occurs with a rate $\hat{\xi} (\mathbf{r})$ with dimension $m^{-3} s^{-1}$. The spatial dependence of the reaction rate originates for a possibly spatially-varying concentration of catalytic material. Under the assumptions of our model, the influence of the reactant is entirely captured by specifying the function $\hat{\xi} (\mathbf{r})$. As such, we will no longer mention the reactant concentration. To match common terminology, we will henceforth refer to the product as ``solute''. 

The mixture is modelled as a Newtonian, incompressible fluid. Due to the small scales of the pore, the Reynolds number is much smaller than one, and so we describe the fluid flow via the Stokes equation
\begin{equation}\label{eq:app-stokes1}
\eta\nabla^{2}\bm{v}(\mathbf{r},t)      = \nabla P(\mathbf{r},t),
\end{equation}
together with the condition for incompressibility
\begin{equation}\label{eq:app-stokes2}
\nabla\cdot \bm{v}(\mathbf{r},t) = 0,
\end{equation}
where $\bm{v}(\mathbf{r},t) $ is the velocity profile, $\eta$ is the dynamic viscosity, and $P(\mathbf{r},t)$ is the pressure. The solute (and therefore, the mixture) interacts with the wall with an effective potential
\begin{equation}
W(\mathbf{r})=\begin{cases}
U_{wall}(\mathbf{r}), & |\mathbf{r}| \leq R(z)\,,\\
\infty, & \text{otherwise},
\end{cases}\label{eq:pot}
\end{equation}
that accounts for the soft interaction with the pore walls, $U_{wall}(\mathbf{r})$, as well as for the confining of the solute inside the pore. Thus, an inhomogeneous concentration of solute leads to an inhomogeneous body force $-\rho \nabla U_{wall}$ acting on the mixture. At steady state, a pressure gradient develops along the wall, with a corresponding flow that travels up/down the gradient of solute concentration if the potential is repulsive/attractive. This is the phenomenon of self-diffusioosmosis \cite{Anderson1989}.\\
Self-diffusioosmosis is operational within a thin layer, of thickness comparable to the potential range $r_\lambda$, near to the channel walls. Accordingly, for what concerns the velocity profile at positions $r\ll R(z)-r_\lambda$ far away from the walls, we can approximate the contribution of self-diffusioosmosis as an effective slip velocity located at the channel walls\cite{Anderson1989}. We will see that only the longitudinal component of this slip velocity will be needed, and this component is given by
\begin{equation}
\label{eq:vslip}
v_{0}(\mathbf{r} \in \mathbf{r}_w,t)=-\frac{\mathcal{L}}{\beta\eta}\nabla_{||}\rho(\mathbf{r} \in \mathbf{r}_w,t) \cdot \textbf{e}_z\,,
\end{equation}
where $\mathbf{r}_w$ is the set of points defining the channel wall, and $\beta = 1/(k_BT)$ is the inverse thermal energy. Furthermore, $\nabla_{||}$ is the derivative along the surface evaluated at the wall, and $\textbf{e}_z$ is the unit vector in the z direction. $\mathcal{L}$ is the phoretic mobility \cite{Anderson1989} and is given by
\begin{equation}
\curlyL = \int\limits_0^{\infty} r^{\prime}\left\{\exp\left[ -\beta U_{wall}(r^{\prime}) \right] -1  \right\} dr^{\prime}, \label{eq:curlyL}
\end{equation}
where $r^{\prime}$ is the distance to the wall. This expression for $\curlyL$ assumes a potential $U_{wall}$ that depends only on the distance to the wall, and a radius of curvature of the wall that is much larger than the potential range\cite{Anderson1989}. Additionally, the solute concentration is assumed to be in thermal equilibrium within this thin layer where $U_{wall} \neq 0$ \cite{Anderson1989}.
From Eq. \eqref{eq:curlyL}, one may see that $\curlyL$ may be positive or negative, depending on whether the interaction potential between solute molecules and the pore wall is attractive or repulsive, respectively. Finally, periodic boundary conditions are used in the pore inlet ($z=-L$) and outlet ($z=L$). We focus on periodic solutions for two reasons. On the one hand they allow us to keep the model simple. On the other hand, they allow us to characterize the intrinsic dynamics of the system i.e,. far away from additional physical boundaries along the longitudinal direction. These boundary conditions allow us to recover the spontaneous symmetry breaking observed in the simulations of Ref.~[\citenum{Antunes2022}]. 

From Eqs. \eqref{eq:app-stokes1} and \eqref{eq:vslip}, we see that the steady state flows are entirely determined by the concentration of solute. This concentration evolves in time according to
\begin{equation}
\dot{\rho}(\mathbf{r},t)=-\nabla\cdot \textbf{j}(\mathbf{r},t)+\hat{\xi}(\mathbf{r})-\chi\rho(\mathbf{r},t)\,,\label{eq:adv-diff}
\end{equation}
where $\textbf{j}(\mathbf{r},t)$ is the flux of solute, $\hat{\xi}(\mathbf{r})$ is a source term, and $\chi \rho(\mathbf{r},t)$ is a sink term. As discussed before, solute production occurs at the pore wall ($x^2 + y^2 = R^2(z)$), and so $\hat{\xi}(\mathbf{r})$ is represented with a Dirac delta
\begin{equation}
\hat{\xi}(\mathbf{r}) = \xi_S(z) \delta\left(R(z) - \sqrt{x^2 + y^2} \right),
\end{equation}
where $\xi_S(z)$ has dimensions $m^{-2} s^{-1}$, and encodes for the inhomogeneous catalytic coating along the pore's axis of symmetry. It is further assumed to be $2L$-periodic. For what concerns the flux of solute, it contains three contributions: one coming from diffusion, one coming from advection due to fluid flow, and one due to the interaction with the pore walls. Thus, the flux $\textbf{j}(\mathbf{r},t)$ is given by
\begin{equation}
\textbf{j}(\mathbf{r},t) = -D\nabla\rho(\mathbf{r},t) -\beta D\rho(\mathbf{r},t)\nabla W(\mathbf{r}) + \bm{v}(\mathbf{r},t)\rho(\mathbf{r},t),
\label{eq:app-J-0}
\end{equation}
where $D$ is the diffusion coefficient. Owing to the confinement given by the potential (Eq. \eqref{eq:pot}), there is no flux through the pore walls. At the pore inlet ($z=-L$) and outlet ($z=L$), the concentration of solute obeys periodic boundary conditions.
The governing equations presented above are not analytically solvable. In order to get insight into the physical mechanisms responsible for the onset of the diverse scenarios that we have observed numerically~\cite{Antunes2022}, we aim at a reduced model that provides a deeper understanding of the mechanisms involved in active pumping. It will also enable a much faster and simpler computation of the estimated pumping rates. 

We proceed by reducing Eqs.~\eqref{eq:app-stokes1} and~\eqref{eq:adv-diff} to effective 1D equations in a procedure analogous to the Fick-Jacobs approach for diffusion in corrugated pores \cite{Zwanzig1992, Reguera2001, Malgaretti2013}. A graphical sketch of the following derivation and its assumptions can be found in Appendix \ref{sec:diagram}.

\section{Reduction to effective 1D equations}

\label{sec:FJ}

Equation~\eqref{eq:adv-diff} is a $3D$ partial differential equation whose analytical solution is not easy to find. Moreover, here we are interested in the transport properties of the pore along its axis of symmetry. Therefore we aim at projecting Eq.~\eqref{eq:adv-diff} onto this axis (the $z$ direction). The procedure that we aim at is based on a length scale separation between the (long) longitudinal length scale and the (short) transverse length scale. As such, the theory we now derive is valid for narrow pores ($R_0 \ll L$). Such an approach (called Fick-Jacobs approximation~\cite{Zwanzig1992,Reguera2001,Kalinay2005,Kalinay2008,Martens2011,Dagdug2013,Malgaretti2023}) has been widely used to model the transport of different systems, spanning from colloidal particles~\cite{Reguera2006,Reguera2012,Marconi2015,Malgaretti2016_entropy,Puertas2018} to polymers~\cite{Bianco2016,Locatelli2023} and including electrolytes~\cite{Malgaretti2014,Malgaretti2015,Chinappi2018,Malgaretti2019_JCP} and active systems~\cite{Malgaretti2017,Dagdug2014,Kalinay2022,Antunes2022}.

\subsection{Equation for solute concentration} 
\label{sec:FJ_rho}

In this section, we derive an effective 1D equation for the solute concentration. This equation governs the time evolution of the solute concentration integrated along the pore cross-section. As such, the dynamics in the transverse degrees of freedom is integrated over and does not need to be explicitly solved for. To do so, we will exploit the length scale separation between the potential range $r_{\lambda}$, the average pore radius $R_0$, and the pore length $L$. 

As discussed, the range $r_\lambda$ of the interaction potential between the solute and the wall is typically much smaller than the pore dimensions $R_0$ and $L$, and so we focus on the regime in which the advective and diffusive contributions $ \mathbf{v} \rho$ and  $D\nabla \rho$ in Eq.~\eqref{eq:app-J-0} dominate the term $\beta D\rho(\mathbf{r},t) \nabla W(\mathbf{r})$ in Eq.~\eqref{eq:app-J-0}. 
Accordingly, Eq.~\eqref{eq:app-J-0} reduces to 
\begin{align}
\textbf{j}(\mathbf{r},t) = &-D\nabla\rho(\mathbf{r},t) + \bm{v}(\mathbf{r},t)\rho(\mathbf{r},t).
\label{eq:app-J}
\end{align}
Due to the axial symmetry of the pore, it is convenient to switch to cylindrical coordinates. Let $\mathbf{r}=(x,y,0)$ be a vector running from the axis of symmetry $(x,y) = (0,0)$ to a generic point inside the pore. This vector is perpendicular to the axis of symmetry, and its magnitude $r$ is the distance from the aforementioned point to the axis of symmetry. We also define $\phi$ as the angle formed by \textbf{r} with the xz-plane.\\
Integrating Eq.~\eqref{eq:adv-diff} over the cross-section and exploiting the axial symmetry of the system leads to
\begin{align}
&\int\limits_{0}^{\infty} \dot{\rho}(r,z,t) r dr =  \nonumber \\
&=- \int\limits_{0}^{\infty} \partial_z j_z(r,z,t) rdr +  \int_{0}^{\infty}[\hat{\xi}(r,z)-\chi\rho(r,z,t)] r dr.
\label{eq:adv-diff2}
\end{align}
The above step exploits the fact that, due to Eq.~\eqref{eq:pot}, there is no solute anywhere outside the pore, i.e.,
\begin{align}
    \rho \Big(r > R(z) ,z,t\Big)=0,
\end{align}
and
\begin{align}
    j_r\Big(r > R(z),z,t\Big)=0\,,
\end{align}
and therefore,
\begin{equation}
    \int\limits_0^\infty \partial_r [r j_r(r,z,t)] dr = 0.
        \label{eq:integratedJ}
\end{equation}
As discussed, we will now focus on the case of narrow pores such that
\begin{equation}
    R_0 \ll L,
\end{equation}
which introduces a separation of length scales between the radius and the length of the channel. The result is also a separation between the relaxation timescale in the radial and in the longitudinal direction, with the relaxation of $\rho$ occurring much faster in the former. Thus, we employ the Fick-Jacobs approximation \cite{Zwanzig1992, Reguera2001, Malgaretti2013}, which consists in assuming no net transport in the radial direction, i.e.
\begin{equation}
\label{eq:fickjacobs}
   j_r(r,z,t) = 0\,. 
\end{equation}
As a result of this fast relaxation, we may examine the dynamics in the radial direction separately from the longitudinal direction. We begin by estimating the magnitude of the advective transport via the incompressibility equation (Eq. \eqref{eq:app-stokes2}), i.e.,
\begin{align}
    \nabla\cdot\bm{v}(\mathbf{r},t)=\frac{1}{r}\partial_r\left(r v_r(\mathbf{r},t)\right)+\partial_z v_z(\mathbf{r},t)= 0\,,
    \label{eq:app-incompr}
\end{align}
where in the last step we have assumed that the axial symmetry of the pore leads to an axial symmetry of the velocity profile, i.e., $\partial_\phi \bm{v}=0$. We now estimate how large the velocities in the radial direction are, when compared to those in the longitudinal direction. We introduce dimensionless variables
\begin{align}
 \hat{r} &= \frac{r}{R_0}, \label{eq:lubDef0} \\
 \hat{z} &= \frac{z}{L},\\
 \hat{v}_z(\hat{r}, \hat{z}, t) &= \frac{v_z(\hat{r}, \hat{z}, t)}{v^*_z},\\
 \hat{v}_r(\hat{r}, \hat{z}, t) &= \frac{v_r(\hat{r}, \hat{z}, t)}{v^*_r},\label{eq:lubDef3}
\end{align}
where $v^*_z$ and $v^*_r$ are the order of magnitude of the longitudinal and transverse components of the velocity, respectively. Thus, hatted variables are dimensionless and take values of order unity. Using these definitions, we may write Eq. \eqref{eq:app-incompr} as 
\begin{equation}
\label{eq:radialv}
    \frac{v^*_r}{v^*_z} = \frac{R_0}{L} K,
\end{equation}
where $K$ is a quantity of order unity given by
\begin{equation}
    K = \partial_{\hat{z}} \hat{v}_z (\hat{r}, \hat{z}, t) \left[ \hat{r}^{-1} \partial_{\hat{r}} (\hat{r} \hat{v}_r(\hat{r}, \hat{z},  t) \right]^{-1}.
\end{equation}
Due to the form of Eq. \eqref{eq:radialv}, $K$ must be a constant.

We now introduce the radial P\'eclet number $Pe_r$ by quantifying the relative magnitude of diffusive and advective timescales in the radial direction:
\begin{equation}
    Pe_r = \frac{R_0 v^*_r}{D} ,
\end{equation}
and as per Eq.~\eqref{eq:radialv}
\begin{align}
    Pe_r= \frac{R^2_0}{L}\frac{v_z^*}{D} K.
\end{align}
In a sufficiently narrow pore ($R_0 \ll L)$, one has $Pe_r \ll 1$. This means that diffusive transport dominates the advection due to the fluid flow in the radial direction. 

For pores that are narrow enough, diffusion leads to a solute concentration that depends weakly on the radial direction, despite solute being produced only at the pore wall. Furthermore, the flow field is entirely determined by the solute concentration at the wall alone (via the slip boundary condition of Eq. \eqref{eq:vslip}). We make use of this property, and neglect the variation of solute concentration in the radial direction:
\begin{equation}
\label{eq:ansatzIsotropic}
\rho(r,z,t) \equiv \rho(z,t) \ = \ \frac{p(z,t)}{\pi R^2(z)},
\end{equation}
Accordingly, inserting Eq.~\eqref{eq:ansatzIsotropic} into Eq.~\eqref{eq:adv-diff2} and using Eq.~\eqref{eq:app-J} leads to
\begin{align}
\pi R^2(z)\dot{\rho}(z,t)=&D\partial_z^2[\pi R^2(z) \rho(z,t)]  - \partial_z\Big[Q(t)\rho(z,t)\Big] + \nonumber \\
             &+ 2\pi R(z) \xi_S(z) - \pi R^2(z)  \chi \rho(z,t),\label{eq:toExpand_st0}
\end{align}
with
\begin{equation}
Q(t) = 2 \pi \int\limits_{0}^{R(z)} v_z(r,z,t)r dr.
\label{eq:Q_comp}
\end{equation}
We note that $Q$ is the volumetric fluid flow which, due to incompressibility (see Eq.~\eqref{eq:app-stokes2}), does not vary along the $z$ direction. 
For later use, we define the integrated source $\xi(z)$ and integrated solute concentration $p(z,t)$
\begin{align}
     \xi(z) &= 2\pi R(z) \xi_S(z) ,\\
     p(z,t) &= \pi R^2(z) \rho(z,t) ,
\end{align} 
which, when plugged-in to Eq.~\eqref{eq:toExpand_st0} yields
\begin{align}
\dot{p}(z,t)=&D\partial_z^2p(z,t)  - \partial_z\Big[Q(t)\frac{p(z,t)}{\pi R^2(z)}\Big] + \nonumber \\
             &+ \xi(z) - \chi p(z,t).\label{eq:toExpand_st}
\end{align}

\subsection{Equation for flow rate} 

\label{sec:FJ_vel}

In this section, we handle the Stokes equation to obtain the flow rate $Q$, as required for the solution of Eq. \eqref{eq:toExpand_st}. To do so, we once again make use of the length separation between the average pore radius $R_0$ and the pore length $L$. Such a separation leads to the velocity in the longitudinal direction varying strongly in the radial direction, as compared with the longitudinal direction. Further requesting that the pore radius varies smoothly (($\partial_z R(z))^2 \ll 1 $) enables $Q$ to be written as a functional of the solute concentration.\\

As discussed, to solve Eq. \eqref{eq:toExpand_st}, one has to express $Q(t)$ in terms of $p(z,t)$. As mentioned in section \ref{sec:FJ_rho}, the pore is considered to be narrow ($L \gg R_0$) and axially symmetric, as is the velocity profile. Within such a regime, we exploit the lubrication approximation \cite{Schlichting1979}. To do so, we write Eq. \eqref{eq:app-stokes1} for both longitudinal and transverse coordinates
\begin{align} 
    &\eta \left\{ \partial^2_z v_z(r,z,t) + r^{-1} \partial_r \left[ r \partial_r v_z(r,z,t) \right]  \right\} = \partial_z P(r, z,t), \label{eq:stokesLub0}\\
    &\eta \left\{  \partial^2_z v_r(r,z,t)  + r^{-1} \partial_r \left[r \partial_r v_r(r,z,t)\right] - r^{-2} v_r(r,z,t) \right\} = \nonumber\\
    &= \partial_r P(r,z,t), \label{eq:stokesLub1}
\end{align}
and introduce the dimensionless pressure
\begin{align}
 \hat{P}(\hat{r}, \hat{z}, t) = \frac{P(r,z,t)}{P^*},
\end{align}
where $P^*$ is the order of magnitude of the pressure. With this definition and those of Eqs. \eqref{eq:lubDef0} - \eqref{eq:lubDef3} , Eq. \eqref{eq:stokesLub0} can be written as
\begin{equation}
   \frac{\eta  v^*_z L }{P^* R_0^2}\left\{ \frac{R_0^2}{L^2} \partial^2_{\hat{z}} \hat{v}_z(\hat{r},\hat{z},t) + \hat{r}^{-1} \partial_{\hat{r}} \left[ \hat{r} \partial_{\hat{r}} \hat{v}_z(\hat{r},\hat{z},t) \right]  \right\} = \partial_{\hat{z}} \hat{P}(\hat{r}, \hat{z},t),  \label{eq:stokesLub2} 
\end{equation}
and since the pore is narrow ($R_0 \ll L$), the left hand side of Eq. \eqref{eq:stokesLub2} may be approximated as
\begin{equation}
          \frac{\eta  v^*_z L }{P^*R_0^2} \hat{r}^{-1} \partial_{\hat{r}} \left[ \hat{r} \partial_{\hat{r}} \hat{v}_z(\hat{r},\hat{z},t)  \right]= \partial_{\hat{z}} \hat{P}(\hat{r}, \hat{z},t).  \label{eq:stokesLub3} 
\end{equation}
We now write Eq. \eqref{eq:stokesLub1} as
\begin{align}
      & \frac{\eta  v^*_z K }{P^*L}\left\{ \frac{R_0^2}{L^2} \partial^2_{\hat{z}} \hat{v}_r(\hat{r},\hat{z},t) + \hat{r}^{-1} \partial_{\hat{r}} \left[ \hat{r} \partial_{\hat{r}} \hat{v}_r(\hat{r},\hat{z},t) \right] - \hat{r}^{-2} \hat{v}_r  \right\} = \nonumber \\
       &=\partial_{\hat{r}} \hat{P}(\hat{r}, \hat{z},t), \label{eq:stokesLub4}
\end{align}
where we have used Eq. \eqref{eq:radialv}. By comparing Eqs. \eqref{eq:stokesLub2} and \eqref{eq:stokesLub4}, we may write
\begin{equation}
    \partial_{\hat{r}} \hat{P}(\hat{r}, \hat{z},t)\left[ \partial_{\hat{z}} \hat{P}(\hat{r}, \hat{z},t) \right]^{-1} = \mathcal{O} \left( \frac{R_0^2}{L^2} \right),
\end{equation}
and thus the variation of the pressure along the transverse direction is negligible
\begin{equation}
    P(r,z,t) \equiv P(z,t).
\end{equation}
As a result, we may write Eq. \eqref{eq:stokesLub3} as
\begin{equation}
     \eta r^{-1} \partial_r \left[ r \partial_r v_z(r,z,t) \right] =  \partial_z P(z,t),  \label{eq:stokesLub5}
\end{equation}
where we have restored the dimensional quantities. Integrating Eq. \eqref{eq:stokesLub5} twice along the radial direction  leads to
\begin{equation}
v_z(r,z,t) = v_0(z,t) - \frac{\partial_z P(z,t)}{4 \eta} \Big[ R^2(z) - r^2 \Big]\,,
\label{eq:plugPlusPoi}
\end{equation}
where we have used the diffusioosmotic slip boundary conditions $v_z(r=R,z,t) = v_0(z,t)$. Note that $\partial_{z}P(z,t)$ contains, in addition to a possible external pressure drop, also contributions stemming from fluid incompressibility that will ensure that $Q$ does not depend on $z$. 
The volumetric fluid flow is then
\begin{align}
Q(t)=v_0(z,t) \pi R^2(z)-\frac{\pi}{8}\frac{\partial_z P(z,t)}{ \eta}R^4(z)\,.
\end{align}
 As discussed, we focus on cases where $P(z)$ fulfills periodic boundary conditions. As such, the integral of $\partial_z P(z,t)$ over the pore length has to vanish, i.e.
\begin{align}
    0=&\int\limits_{-L}^L \partial_z P(z,t) dz = 8 \pi \int\limits_{-L}^L \frac{v_0(z,t)}{R^{2}(z)} dz -\nonumber \\
    &- \frac{8 \eta}{\pi}Q(t) \int\limits_{-L}^L\frac{dz}{R^{4}(z)}.
\end{align}
This allows one to determine $Q(t)$:
\begin{equation}
Q(t)  = \pi \int\limits_{-L}^ L \dfrac{v_0(z,t)}{R^2(z)} dz  \Big/  \int\limits_{-L}^ L \dfrac{dz}{R^4(z)} \,.
\label{eq:Q}
\end{equation}
Finally, we need to write the slip velocity $v_0(z,t)$ as a function of $p(z,t)$. The vector perpendicular to the pore wall is:
\begin{equation}
    \textbf{n}(\phi,z) = \left[1 + \Big(\partial_z R(z)\Big)^2\right]^{-1/2}(- \textbf{e}_r + \partial_z R(z) \textbf{e}_z)\,,
\end{equation}
where $\textbf{e}_r$ is the unit vector pointing in the radial direction. With this expression Eq.~\eqref{eq:vslip} turns into
\begin{align}
&v_{0}(z,t) = \nonumber \\
&= -\frac{\mathcal{L}}{\beta\eta} \Big[ \nabla\rho(r=R,t) \cdot \bm{e}_z - (\nabla\rho(r=R,t) \cdot \bm{n}) \bm{n} \cdot \bm{e}_z \Big] = \nonumber \\
& = -\frac{\mathcal{L}}{\beta\eta}\left[ \partial_z\rho-\frac{(\partial_z \rho) (\partial_z R) + r^{-1}\partial_r(r\partial_r \rho)}{\sqrt{1 + (\partial_z R)^2}} \cdot \frac{\partial_z R}{\sqrt{1 + (\partial_z R)^2}}\right].  \label{eq:v0toapproximate}
\end{align}
We recall that Eq.~\eqref{eq:app-J} is particularly valid in the regime of weakly varying pore radii. In this regime one has $ (\partial_z R)^2 \ll 1 $, and Eq.~\eqref{eq:v0toapproximate} can be approximated as
\begin{align}
\label{eq:v02}
v_{0}(z,t) \approx -\frac{\mathcal{L}}{\beta\eta}\partial_z\rho (z,r=R,t)  \ =  -\frac{\mathcal{L}}{\beta\eta}\partial_z\left[\frac{p(z,t)}{\pi R^2(z)} \right],
\end{align}
where we have used $\partial_r \rho =0$ coming from Eq. \eqref{eq:ansatzIsotropic}. By using Eqs.~\eqref{eq:ansatzIsotropic} and~\eqref{eq:v02}, Eq.~\eqref{eq:Q} becomes
\begin{align}
\label{eq:Q_expanded}
Q(t)  =& -\frac{\mathcal{L}}{\beta \eta} \left[ \int\limits_{-L}^ L \dfrac{dz}{R^4(z)} \right]^{-1} \int\limits_{-L}^ L R^{-2}(z) \partial_z[ R^{-2}(z) p(z,t) ]  dz.
\end{align}

\subsection{The case of weakly-corrugated pores} 

\label{sec:FJ_deltaR}

The 1D integro-differential equation obtained by plugging Eq. \eqref{eq:Q_expanded} in Eq. \eqref{eq:toExpand_st} remains challenging to solve analytically. Accordingly, we will expand both equations to linear order in the corrugation height. This truncation is a good approximation for weakly-corrugated pores, and allows for an analytical solution for the steady state. We thus define
\begin{eqnarray}
R(z) = R_0 + \delta R(z), 
\end{eqnarray} 
where
\begin{equation}
\INTL \delta R(z) = 0,
\end{equation}
where in the limit of low corrugation
\begin{equation}
\delta R(z) \ll R_0.
\end{equation}
We also expand $p(z,t)$ in $\delta R(z)$,
\begin{align}
\label{eq:expansionInDeltaR_p}
p (z,t;\delta R(z)) = p_0(z,t) + \sum\limits_{l>0} p_{l}(z,t) \delta R^l(z).
\end{align}
From Eqs.\eqref {eq:toExpand_st} and~\eqref{eq:Q_expanded} we obtain
\begin{widetext}
\begin{align}
\label{eq:Q_expanded2}
&Q(t)  = -\frac{\mathcal{L}}{\beta \eta} \left[ \int\limits_{-L}^ L \dfrac{dz}{[R_0 + \delta R(z)]^4(z)} \right]^{-1}  \int\limits_{-L}^ L [ R_0 + \delta R(z)]^{-2} \partial_z \left\{ [R_0 + \delta R(z)]^{-2} \left( p_0(z,t) + \sum\limits_{l>0} p_l(z,t) \delta R^l (z) \right) \right\}  dz,\\
&\dot{p}_0(z,t) + \sum\limits_{l>0} \dot{p}_l(z,t) = D \partial_z^2 \left( p_0(z,t) + \sum\limits_{l>0} p_l(z,t) \delta R^l (z) \right)  - \frac{Q(t)}{\pi} \partial_z \left\{ [R_0 + \delta R(z)]^{-2} \left[p_0(z,t) + \sum\limits_{l>0} p_l(z,t) \delta R^l (z) \right] \right\} + \nonumber \\
& + 2\pi [R_0 + \delta R(z)] \xi_S(z) - \chi \left(p_0(z,t) + \sum\limits_{l>0} p_l(z,t) \delta R^l (z) \right). \label{eq:p_expanded2}
\end{align}
\end{widetext}
Using the fact that both $R(z)$ and $p(z)$ are periodic, the leading term in the right-hand-side of Eq.~\eqref{eq:Q_expanded2} is
\begin{equation}
\label{eq:Qintegral}
Q(t)  = \frac{\mathcal{L}}{\beta \eta} \frac{R_0}{L} \int\limits_{-L}^ L \delta R(z)  \partial_z p_0(z,t)  dz. 
\end{equation}
From Eq.~\eqref{eq:p_expanded2}, we find the contribution $p_0(z)$ as the solution of 
\begin{equation}
\label{eq:eqForP0}
\dot{p}_0(z,t) = D\partial_z^2 p_0 (z,t) - \chi p_0(z,t) + 2\pi R_0 \xi_S(z) - \frac{Q(t)}{\pi R_0^2} \partial_z p_0(z,t).
\end{equation}
We will henceforth neglect contributions to $p(z,t)$ of higher order in $\delta R$. As the values $p_l$ for $l>0$ will no longer be considered, we simplify the notation by dropping the subscript and renaming $p_0(z,t)$ as $\mathcal{P}(z,t)$
\begin{equation}
\mathcal{P}(z,t) \equiv p_0(z,t).
\end{equation}
We now perform a Fourier expansion in space of $\mathcal{P}(z,t)$ and $\xi_S(z)$, such that
\begin{equation}
\mathcal{P}(z,t)=\mathcal{P}_0(t) + \sum\limits_{j>0} \mathcal{P}_j(t) \cos\left( k_j z  \right) +  \sum\limits_{j>0} \tilde{\mathcal{P}}_j(t) \sin\left( k_j z  \right), \label{eq:def-p2}
\end{equation}
\begin{equation}
2\pi R_0 \xi_S(z)=\xi_0 + \sum\limits_{j>0} \xi_j \cos\left( k_j z  \right) + \sum\limits_{j>0} \tilde{\xi}_j \sin\left( k_j z \right)\label{eq:def-xi2},
\end{equation}
where we define the Fourier coefficients $\{\mathcal{P}_j\}$ and $\{\mathcal{\xi}_j\}$ associated to the symmetric part of the expansion, and the coefficients $\{\tilde{\mathcal{P}}_j\}$ and $\{\tilde{\mathcal{\xi}}_j \}$ associated to the antisymmetric part of the expansion. We have also defined $k_j = (\pi/L) j$. Plugging Eqs.~\eqref{eq:def-p2} and~\eqref{eq:def-xi2} in Eq.~\eqref{eq:eqForP0} leads to
\begin{align}
\dot{\mathcal{P}}_j(t) &= a_j  \mathcal{P}_j(t) + Qb_j\tilde{\mathcal{P}}_j(t) + \xi_j, \label{eq:system_p}\\ 
\dot{\tilde{\mathcal{P}}}_j(t) &= a_j  \tilde{\mathcal{P}}_j(t) - Q(t)b_j \mathcal{P}_j(t)  +\tilde{\xi}_j,\label{eq:system_pTilde}
\end{align}
where
\begin{align}
\label{eq:a_def}
a_j &= - \left[ \chi + D k_j^2 \right], \\
\label{eq:b_def}
b_j &= - \frac{k_j}{\pi R_0^2}.
\end{align}
We insert Eq.~\eqref{eq:def-p2} into Eq.~\eqref{eq:Qintegral} to obtain
\begin{equation}
\label{eq:centralQ}
Q(t) = - \frac{\mathcal{L}}{\beta\eta } \sum\limits_{j=1}^\infty ( \tilde{\mathcal{P}}_j(t) \Gamma_j  - \mathcal{P}_j(t) \Theta_j ),
\end{equation}
where
\begin{align}
\Gamma_j =& - \frac{ k_j}{LR_0}  \left[\int\limits_{-L}^L \delta R (z) \cos\left( k_j z \right) dz \right],\label{eq:defGamma}\\
    \Theta_j = &- \frac{k_j}{LR_0}  \left[\int\limits_{-L}^L \delta R (z) \sin\left( k_j z \right) dz \right],\label{eq:defTheta}
\end{align} 
which depends only on the geometrical properties of the pore. If we write
\begin{equation}
\delta R(z) = \sum_{j=1}^\infty R_{j}\cos(k_j z) + \sum_{j=1}^\infty \tilde{R}_{j}\sin(k_j z) ,
\end{equation}
we obtain
\begin{align}
\Gamma_j =& - \frac{ k_j}{R_0}  R_j,\label{eq:defGamma2}\\
    \Theta_j = &- \frac{k_j}{R_0}  \tilde{R}_j.
\end{align}

In steady-state, Eqs.~\eqref{eq:system_p} and~\eqref{eq:system_pTilde} result in
\begin{eqnarray}
a_j  \mathcal{P}_j + Qb_j\tilde{\mathcal{P}}_j  + \xi_j = 0,\\
a_j  \tilde{\mathcal{P}}_j - Qb_j \mathcal{P}_j  + \tilde{\xi}_j = 0,
\end{eqnarray}
which, if $a_j^2 +b_j^2Q^2 \neq 0$, is solved by 
\begin{eqnarray}
\label{eq:FourierPStart}
\mathcal{P}_j = \frac{Qb_j \tilde{\xi}_j - a_j \xi_j}{a_j^2+b_j^2Q^2},\\
\tilde{\mathcal{P}}_j = -\frac{a_j \tilde{\xi}_j + Q b_j \xi_j}{a_j^2+b_j^2Q^2}.\label{eq:FourierPEnd}
\end{eqnarray}
We note that in the non-pumping case $Q=0$, even/odd contributions to the solute distribution depend solely on the even/odd contributions of the source function $\xi(z)$.

\section{Sinusoidal pore}

\label{sec:sinusoidal}

Equation~\eqref{eq:centralQ} together with Eqs.~\eqref{eq:system_p} and~\eqref{eq:system_pTilde} results in a polynomial equation of (in principle) infinite degree for the flow rate $Q$. While for a general shape $R(z)$ and source $\xi(z)$ finding the roots of such an equation analytically is not possible, the problem becomes much simpler in the case of a sinusoidal pore such that
\begin{equation}
    R(z) = R_0 + R_1 \cos(k_1 z)+ \tilde{R}_1 \sin(k_1 z). \label{eq:sinusoidalshape}
\end{equation}
Without loss of generality, we may place the center of our frame of reference where the pore is narrowest (bottleneck), meaning $\tilde{R}_1 = 0$ and $R_1 <0$. The larger the value of $|R_1|$, the larger the height of the corrugation. Note that our assumption of weak corrugation limits our approach to values of $|R_1|/R_0 \ll 1$. With the definition of the shape $R(z)$ of Eq.~\eqref{eq:sinusoidalshape}, Eq.~\eqref{eq:centralQ} now yields
\begin{equation}
\label{eq:Q_LR}
Q =  \frac{\mathcal{L}}{\beta\eta } \tilde{\mathcal{P}}_1 k_1 \frac{R_1}{R_0} \ , \ R_1 \leq 0.
\end{equation}
The fact that the shape $R(z)$ only contains one Fourier mode leads to the conclusion that only the very same mode of the source function contributes to the flow rate. The reverse is also true: should the source function $\xi(z)$ only contain one Fourier mode, only the same Fourier mode of the shape $R(z)$ contributes to $Q$. Thus, without loss of generality, we may write the source function as
\begin{equation}
\label{eq:source_brokenChem}
\xi(z) = \xi_0 + \xi^{\prime} \cos\left[ k_1 (z-z_0) \right],
\end{equation}
where $\xi^{\prime}>0$ is the amplitude of the variation of the source, and $z_0$ is the shift between the maximum of the source and the minimum of $R(z)$. It is useful to write Eq.~\eqref{eq:source_brokenChem} in the form of 
\begin{equation}
\xi(z) = \xi_0 + \xi_1 \cos(k_1 z)+ \tilde{\xi}_1 \sin(k_1 z),
\end{equation}
where
\begin{align}
    \xi_1 &= \xi^{\prime} \cos( k_1 z_0 ), \label{eq:shiftedXi1}\\
    \tilde{\xi}_1 &= \xi^{\prime} \sin( k_1 z_0 ). \label{eq:shiftedTildeXi1}
\end{align}

\subsection{In-phase pore shape and source ($z_0 = 0$) }

\label{sec:sinusoidalSym}

\begin{figure*}[t]
	\centering
   \includegraphics[width=1.0\textwidth]{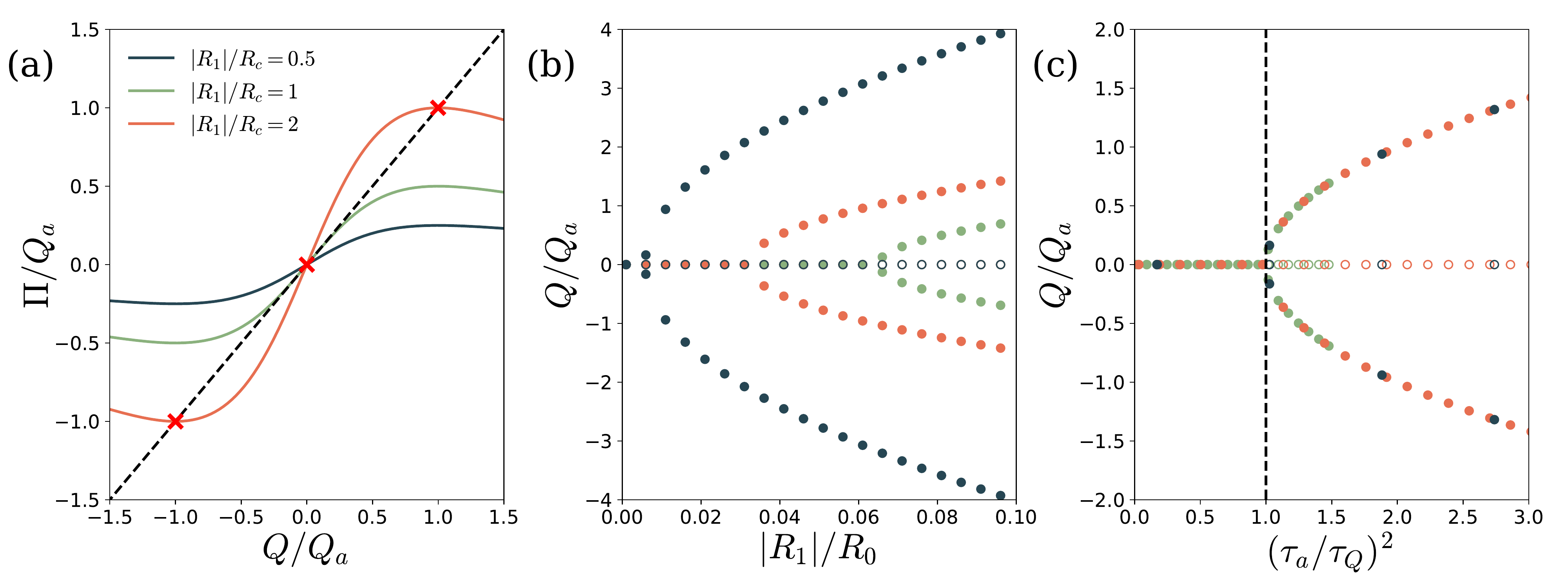}
   \caption{ (a) The emergence of three real solutions, corresponding to the pumping transition. The dashed black line marks the curve $\Pi = Q$. The intersections of that line with the lines obtained from Eq.~\eqref{eq:QOnlyOneXi} are marked with a red cross and correspond to the steady-state values of $Q$. Parameters are $(\mathcal{L}/\beta\eta) (\tau_{\chi}/L^5) = -10^{-6}$, $\xi_0 L \tau_{\chi} = \xi_1 L \tau_{\chi} = 5 \times 10^4$, $R_0/L = 0.1$, $D \tau_{\chi} /L^2 = 10^{-3}$.  (b) The possible steady state flow rates as function of $|R_1|/R_0$ for the previous parameters excluding $R_0/L$. Data shown corresponds to $R_0/L \in \{ 3 \times 10^{-2} \text{ (\textcolor[HTML]{264653}{gray}), } 7 \times 10^{-2} \text{ (\textcolor[HTML]{E76F51}{orange}), } 10^{-1} \text{ (\textcolor[HTML]{8AB17D}{green})}\}$. Open symbols correspond to unstable steady states and filled-in symbols correspond to stable steady states. (c) Collapse of the data in panel (b)  \label{fig:sinusoidal_symmetric_adim} } 
\end{figure*}

\begin{figure}[t]
	\centering
   \includegraphics[width=0.5\textwidth]{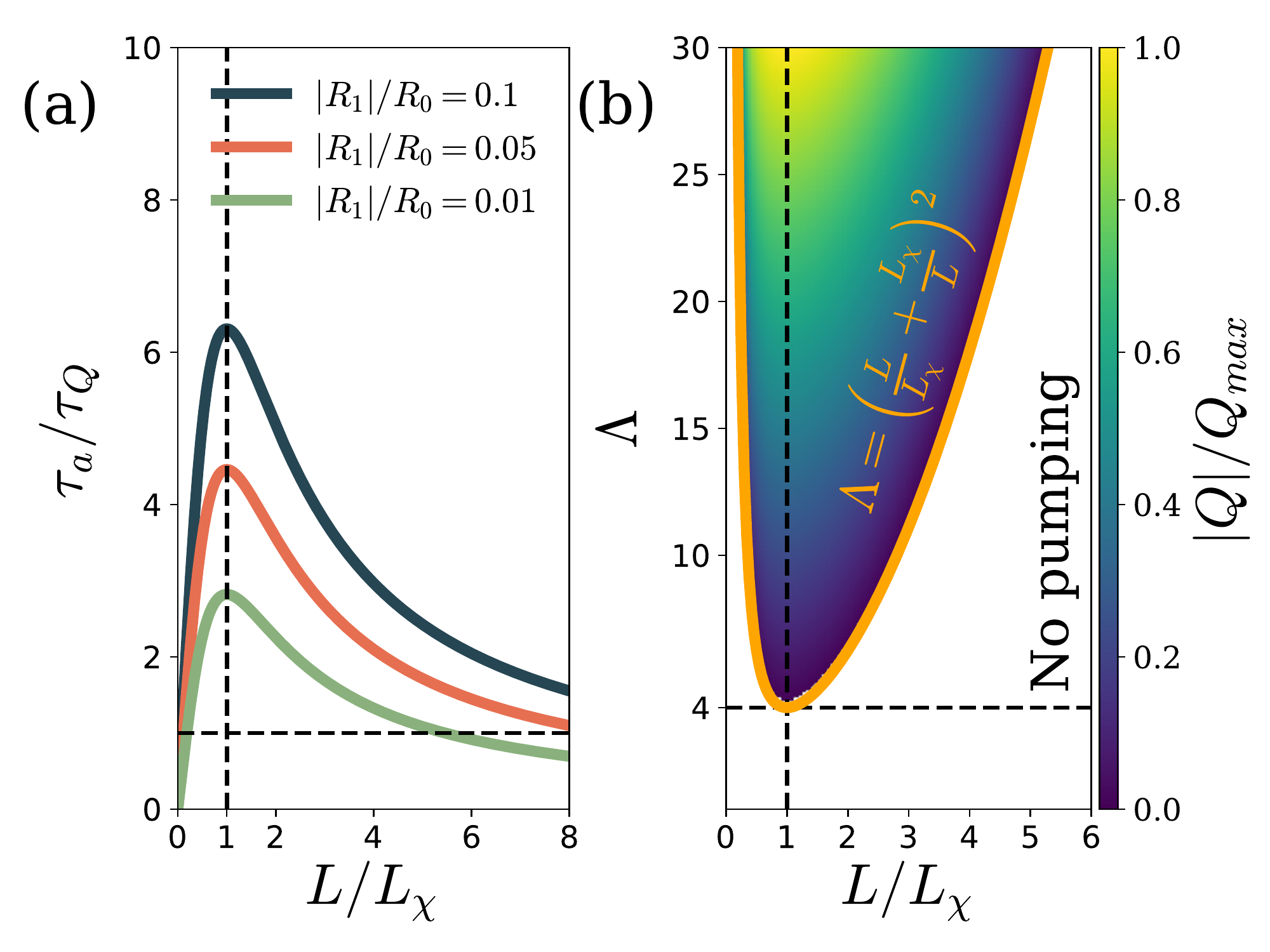}
   \caption{ (a) Variation of the dimensionless parameter $\tau_a/\tau_Q$ with $L/L_{\chi}$ for three values of $|R_1|/R_0$. Parameters are $(\mathcal{L}/\beta\eta) (\tau_{\chi}/L^5) = -10^{-6}$, $\xi_0 L \tau_{\chi} = \xi_1 L \tau_{\chi} = 5 \times 10^4$, $R_0/L = 0.1$. (b) The possible steady state flow rates as function of $L/L_{\chi}$ and $\Lambda$. The flow rates are normalized by the maximum flow rate in the diagram. The horizontal dashed line identifies the minimum value of $\Lambda$ for which pumping still occurs. The vertical dashed line identifies the maximal value of $L/L_{\chi}$. The full line shows the transition between pumping and non-pumping. \label{fig:sinusoidal_symmetric_lenght}} 
\end{figure}

We first investigate the case $z_0=0$, where the pore is fore-aft symmetric, i.e. $R(z) = R(-z)$ and $\xi(z) = \xi(-z)$. In this case, Eq.~\eqref{eq:Q_LR} reduces to
\begin{equation}
\label{eq:QOnlyOneXi}
    Q =  -\frac{\mathcal{L}}{\beta\eta } \left(-k_1 \frac{R_1}{R_0}\right) \frac{-b_1 \xi_1 Q}{a_1^2 +b_1^2Q^2} \equiv \Pi(Q),
\end{equation}
where $\Pi(Q)$ is the polynomial on the right-hand-side of Eq.~\eqref{eq:QOnlyOneXi}. First, we note that both $Q$ and $-Q$ are solutions, as required from the symmetry of the pore. Furthermore, the non-pumping state $Q=0$ is always a solution, albeit not always a stable one. A linear stability analysis reveals that $Q=0$ is only stable if it is the only real solution (see Appendix \ref{sec:LSA}). There is only a non-zero solution if $\Pi(Q)$ is such that
\begin{equation}
    \frac{\partial}{\partial Q}\Pi(Q)|_{Q=0} > 1,
\end{equation}
as seen graphically in Fig.~\ref{fig:sinusoidal_symmetric_adim} (a). This condition yields
\begin{equation}
    a_1^2 < -\frac{\mathcal{L}}{\beta \eta} \left(-k_1 \frac{R_1}{R_0}\right) (-b_1)\xi_1,
\end{equation}
which, combined with Eqs.~\eqref{eq:a_def} and~\eqref{eq:b_def} results in a minimum corrugation height $R_c$ such that pumping only occurs when
\begin{equation}
\label{eq:pumpingCond}
    |R_1| > R_c,
\end{equation}
with the critical value
\begin{equation}
\label{eq:critR}
R_c = -\frac{1}{\pi }\frac{1}{ \curlyL \xi_1}\left( 1 + \frac{\pi^2 D}{L^2 \chi} \right)^2 \beta \eta L^2 \chi^2 R_0^3.
\end{equation}
It is important to note the special case of a flat pore ($R_1 =0$), for which no pumping is possible, as seen in Fig.~\ref{fig:sinusoidal_symmetric_adim} (b). 
The solutions $Q \neq 0$ can be obtained from Eq.~\eqref{eq:QOnlyOneXi} yielding
\begin{align}
Q^2=  \pi \frac{\mathcal{L}}{\beta \eta} R_0 R_1 \xi_1 - \left[ \chi + D\left( \frac{\pi}{L} \right)^2 \right]^2 L^2R_0^4. \label{eq:GetQ}
\end{align}
It must be noted that there can only be pumping ($Q \neq 0$) if
\begin{equation}
    \mathcal{L} \xi_1 < 0.
\end{equation}
This means that we can only obtain pumping in two situations: If the potential is repulsive\footnote{See Eq.~\eqref{eq:curlyL} for the definition of $\curlyL$} ($\mathcal{L} < 0$), then the source has to be strongest at the bottleneck ($\xi_1 > 0$). This result is consistent with previous works on active pores \cite{Antunes2022}. If the potential is attractive ($\mathcal{L} > 0$), the exact opposite has to occur. Pumping then occurs only when the source is strongest where the pore is widest ($\xi_1 <0$). This is analogous to the case of the isotropic diffusiophoretic colloids \cite{Michelin2013,deBuyl2013}. From here on out, for simplicity and clarity of the results, we will assume $\curlyL <0$, $\xi_1>0$, but the generalization is straightforward. 

To understand Eq.~\eqref{eq:GetQ}, we define two timescales. One timescale corresponds to the average time between creation and destruction of a solute molecule, which we call the average lifetime
\begin{align}
    \tau_{\chi} &= \frac{1}{\chi}. \label{eq:tauChi}
\end{align}
The second timescale $\tau_D$ characterizes diffusion. More precisely, it is the time that it takes for a solute molecule to diffuse for a length $L$:
\begin{align}
    \tau_D &= \frac{L^2}{D}. \label{eq:tauD}
\end{align}
Substituting Eqs.~\eqref{eq:tauChi} and~\eqref{eq:tauD} in Eq.~\eqref{eq:GetQ} yields
\begin{align}
\label{eq:GetQ2}
Q^2=  \pi \frac{\mathcal{L}}{\beta \eta} R_0 R_1 \xi_1 - \left[ \frac{1}{\tau_{\chi}} + \frac{\pi^2}{\tau_D}\right]^2 L^2R_0^4.
\end{align}
The smaller either $\tau_{\chi}$ or $\tau_D$ are, the smaller the flow rate will be. Indeed, diffusion acts to inhibit the spontaneous symmetry breaking. The smaller the value of $\tau_D$, the stronger the effect of diffusion. Alongside that process, the timescale $\tau_{\chi}$ limits the length over which a solute molecule may be advected: if a solute molecule does not reside for long enough to reach the bottleneck, the spontaneous symmetry breaking cannot occur either. Thus, we define a new timescale $\tau_a$ which combines these two effects:
\begin{equation}
    \tau_a = \left( \frac{1}{\tau_{\chi}} + \frac{\pi^2}{\tau_D} \right)^{-1}.
\end{equation}
Plugging this quantity into Eq.~\eqref{eq:GetQ2} yields
\begin{align}
\left(\frac{Q}{LR_0^2\tau_a^{-1}}\right)^2=  \frac{ \frac{\pi^2}{L^2}  \frac{\mathcal{L}}{\beta \eta} \frac{R_1}{R_0} \frac{\xi_1}{\pi R_0^2}   - \tau_a^{-2}}{\tau_a^{-2}}.\label{eq:GetQ3} 
\end{align}
We have as of yet not characterized the timescale for advection $\tau_Q$, defined as the time that a solute molecule takes to be advected over a length $L$. Together with $\tau_D$ and $\tau_{\chi}$, $\tau_Q$ characterizes all processes/terms in Eq.~\eqref{eq:adv-diff}. Therefore, it is natural to identify $\tau_Q$ from the terms in Eq.~\eqref{eq:GetQ3} that are not yet accounted for. The shape of Eq.~\eqref{eq:GetQ3} suggests the definition
\begin{equation}
    \tau_Q = \left( \frac{\pi^2}{L^2}  \frac{\mathcal{L}}{\beta \eta} \frac{R_1}{R_0} \frac{\xi_1}{\pi R_0^2} \right)^{-1/2}, \label{eq:tauQ}
\end{equation}
such that one may write
\begin{align}
\left(\frac{Q}{Q_a}\right)^2=  \frac{  \tau_Q^{-2}  - \tau_a^{-2}}{\tau_a^{-2}},\label{eq:GetQ4} 
\end{align}
where we have defined the flow rate scale as
\begin{equation}
    Q_a = \frac{L R_0^2}{\tau_a},
\end{equation}
which roughly corresponds to the flow rate needed to empty half of the pore volume in the time $\tau_a$. Indeed, $\tau_Q$ is the only term that depends on the phoretic mobility $\curlyL$, and therefore on the flow velocities. Equation~\eqref{eq:GetQ4} describes a master curve for $Q$ that is displayed in Fig.~\ref{fig:sinusoidal_symmetric_adim} (c). 
It can be seen from Eq.~\eqref{eq:GetQ4} that the appropriate adimensional number that controls the pumping transition is the ratio $\tau_a/\tau_Q$. In the limit of weak inverse chemical reaction $\tau_{\chi} \gg \tau_D$, the timescale $\tau_a$ is well approximated by $\tau_D$ and
\begin{equation}
   \lim\limits_{\tau_{\chi} \gg \tau_D}  \left(\frac{Q}{Q_a}\right)^2=   \frac{Pe^2}{\pi^4} -1,\label{eq:GetQ5} 
\end{equation}
where $Pe = \tau_D/\tau_Q$ is the P\'eclet number. In the limit of negligible inverse chemical reactions, the pumping condition yields a critical P\'eclet number of $\pi^2$ for all sinusoidal pores (with $z_0 =0)$. 

To further clarify the influence of the pore geometry on the pumping rate, let us examine the role of the pore length $L$ independently of all other parameters. Note that both $\tau_Q$ and $\tau_a$ depend on $L$, but with a different functional form. The timescale $\tau_{\chi}$ is independent of $L$, the timescale $\tau_D$ goes with $L^2$, and the timescale $\tau_Q$ goes with $L$. Thus, the ratio $\tau_a/\tau_Q$ is non-monotonic with $L$, as seen in Fig.~\ref{fig:sinusoidal_symmetric_lenght} (a). This results in two critical values of $L$ such that pumping can only occur if $L_{min} < L < L_{max}$, where
 \begin{align}
\label{eq:maxLOverLchi}
\left(\frac{L_{max}}{L_{\chi}}\right)^2 &= \frac{\Lambda}{2}  -1 + \sqrt{\left(\frac{\Lambda}{2}-1 \right)^2-1},\\
\label{eq:minLOverLchi}
\left(\frac{L_{min}}{L_{\chi}}\right)^2 &= \frac{\Lambda}{2}  -1 - \sqrt{\left(\frac{\Lambda}{2} -1\right)^2-1},
\end{align}
 where
 \begin{equation}
     \Lambda = \frac{\mathcal{L}}{\beta \eta} \frac{R_1}{R_0} \frac{\xi_1}{\pi R_0^2} \frac{1}{\chi D}
 \end{equation}
and 
 \begin{equation}
     L_{\chi} = \pi \sqrt{\frac{D}{\chi}},
 \end{equation}
 which corresponds to the average length a solute molecule diffuses in its lifetime (multiplied by $\pi$). Furthermore, the ratio $\tau_a/\tau_Q$ (and thus $Q/Q_a$) attains a maximum when $L= L_{\chi}$, independently of any other pore parameters, as can be seen in Fig.~\ref{fig:sinusoidal_symmetric_lenght} (b). This diffusive length scale is thus the optimum pore length, no matter the amplitude of the corrugation, the average pore length, or the magnitude of the phoretic mobility.
 
 \subsection{Phase-shift between shape and source ($z_0 \neq 0$)}

\label{sec:sinusoidalAsym}
\begin{figure}[t]
	\centering
   \includegraphics[width=0.5\textwidth]{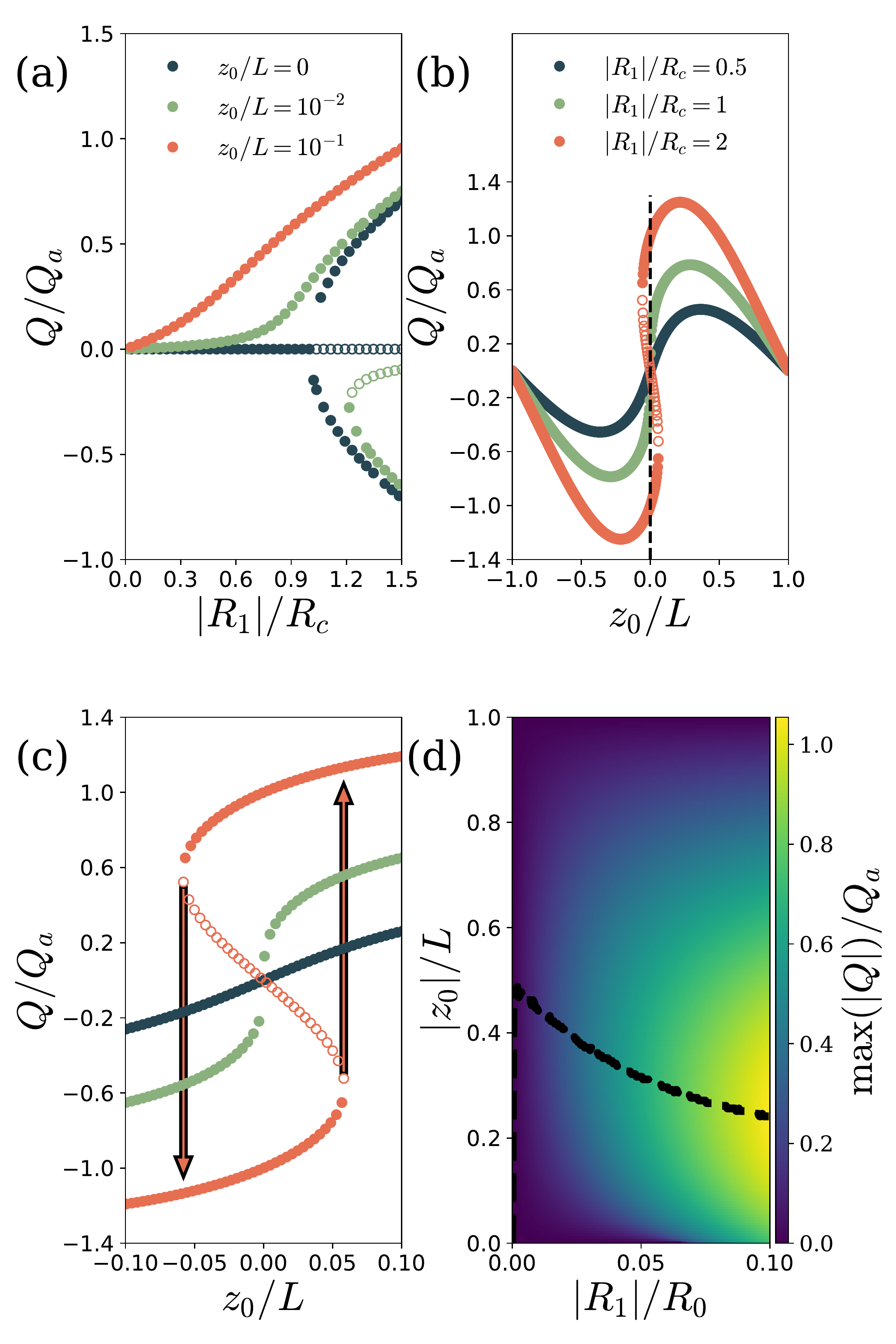}
   \caption{ (a) The possible steady state flow rates as function of $|R_1|/R_c$ for different values of $z_0/L$. Open symbols correspond to unstable steady states and filled-in symbols correspond to stable steady states. Parameters are $(\mathcal{L}/\beta\eta) (\tau_{\chi}/L^5) = -10^{-6}$, $\xi_0 L \tau_{\chi} = \xi_1 L \tau_{\chi} = 5 \times 10^4$, $R_0/L = 0.1$, $D \tau_{\chi} /L^2 = 10^{-3}$.  (b) The possible steady state flow rates as function of $z_0/L$ for different values of $|R_1|/R_c$. Parameters as in (a). (c) Zoom-in of the data in panel (b). Arrows mark the discontinuities that give rise to a hysteresis loop. (d) The maximum flow rate one can obtain as a function of $z_0/L$ and $|R_1|/R_0$. The dotted line marks the value of $z_0/L$ that maximizes $Q/Q_a$ for each value of $|R_1|/R_0$. Parameters as in panel (a).\label{fig:sinusoidal_asymmetric_bifurcation}} 
\end{figure}

\begin{figure}[t]
	\centering
   \includegraphics[width=0.5\textwidth]{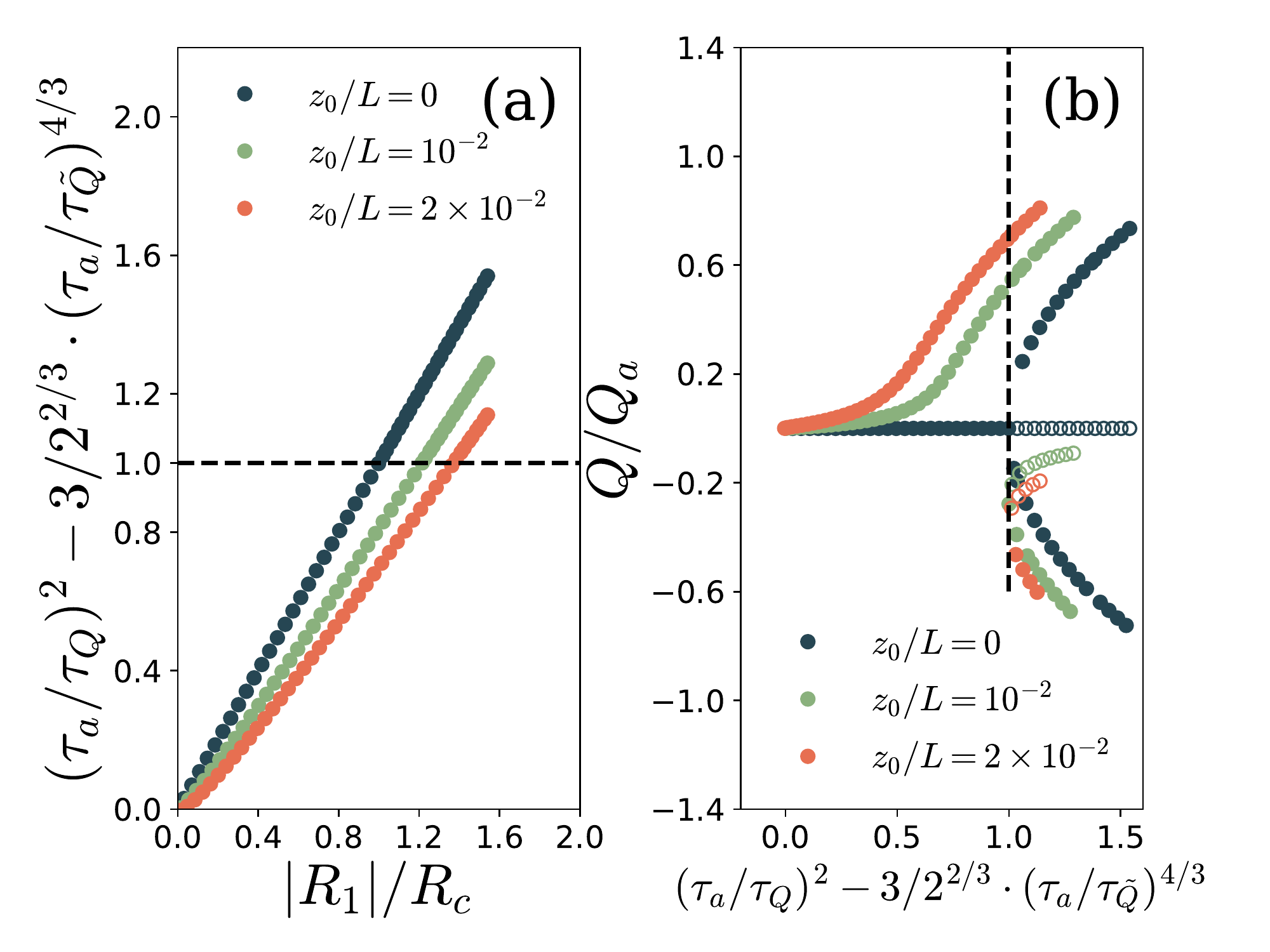}
   \caption{ (a) Variation of the relevant adimensional number that rules the imperfect bifurcation as a function of $|R_1|/R_c$ for three values of $z_0/L$. (b) Pumping rate as a function of the adimensional number. Parameters are $(\mathcal{L}/\beta\eta) (\tau_{\chi}/L^5) = -10^{-6}$, $\xi_0 L \tau_{\chi} = \xi_1 L \tau_{\chi} = 5 \times 10^4$, $R_0/L = 0.1$, $D \tau_{\chi} /L^2 = 10^{-3}$. \label{fig:sinusoidal_asymmetric_adim}} 
\end{figure}

We now break the fore-aft symmetry of the system by introducing an offset $z_0 \neq 0$ between the shape $R(z)$ and the source function $\xi(z)$. Note that both are still sinusoidals, but the bottleneck is no longer located where the source is most intense. We obtain the possible steady-state values of the pumping rate $Q$ from Eq.~\eqref{eq:centralQ} as
\begin{equation}
\label{eq:QBothXi}
        Q =  -\frac{\mathcal{L}}{\beta\eta } \frac{k_1 R_1}{R_0} \frac{a_1 \tilde{\xi}_1+b_1Q\xi_1}{a_1^2 +b_1^2Q^2},
\end{equation}
which has only one real solution if 
\begin{equation}
\left[a_1^2 + \frac{\mathcal{L}}{\beta \eta} \frac{k_1 R_1}{R_0} b_1 \xi_1 \right]^3 > -\frac{27}{4} \left[ \frac{\mathcal{L}}{\beta \eta} \frac{k_1 R_1}{R_0} a_1 b_1 \tilde{\xi}_1 \right]^2.\label{eq:discrimant}
\end{equation}
Note that if $z_0=0$, and $\tilde{\xi}=0$, then Eq.~\eqref{eq:discrimant} reduces to Eq.~\eqref{eq:pumpingCond}. If $z_0 \neq 0$, it is still possible to obtain three real (only two of them stable) solutions of $Q$, meaning pumping in both directions is still possible even when the fore-aft symmetry is broken. The fact that $\tilde{\xi}_1 \neq 0$ means that Eq.~\eqref{eq:QBothXi} cannot be reduced to a quadratic equation, as could Eq.~\eqref{eq:QOnlyOneXi}. As finding the steady state values of $Q$ now requires finding the roots of a cubic polynomial (and the degree increases when considering a non-sinusoidal pore), from here on out, we rely on numerical tools to obtain $Q$. The stability of the steady states is obtained from a linear stability analysis (see Appendix \ref{sec:LSA}).

In Fig.~\ref{fig:sinusoidal_asymmetric_bifurcation} (a), we report the case $z_0=0$ again, as a reference. Slightly shifting the source function ($z_0/L \approx 10^{-2}$), leads to an imperfect bifurcation \cite{Strogatz2015} i.e., a deformation and detachment of the bifurcation branches. A stable branch emerges such that $Q > 0$ for all values of $|R_1|/R_0$. This flow is directed from the bottleneck towards the maximum of the source. A second stable branch with flow going in the opposite direction ($Q < 0$) is still present, but the magnitude of its flow rates, as well as the optimum value of $|R_1|/R_0$ for which it appears, decreases as compared to the symmetric $z_0=0$ case. Note that the two stable branches no longer meet and that the non-pumping state $Q=0$ is no longer a solution. Increasing the offset between source and shape even further ($z_0/L = 10^{-1}$) causes this second stable branch to be pushed to larger values of $|R_1|/R_0$ that are no longer within the regime of validity of this model.
 
It needs to be noted that the values of $Q$ are antisymmetric with $z_0$, as required from the symmetry of the problem. This is clearly seen in Fig.~\ref{fig:sinusoidal_asymmetric_bifurcation} (b). This figure shows also that there can be a range of $z_0/L$ for which two stable branches exist, if the corrugation height is large enough. These two branches also do not meet. Indeed, this implies a discontinuous transition and a sudden inversion of flow if, in an experimental realization, one slowly changes the value of $z_0$ such as to cross from one stable branch to the other. Figure~\ref{fig:sinusoidal_asymmetric_bifurcation} (c) shows how such a procedure leads to a hysteresis loop, with a range of flow rates not accessible to the active pore. 

Conversely, if $z_0/L = \pm 1$, there is no pumping as this corresponds to the $z_0=0$ case discussed before, but with $\xi_1 <0$ i.e., when the minimum value of the source is located in the bottleneck. To characterize the pumping performance of the channel, we now focus on the highest value of $|Q|$ from the three possible solutions, which we call $\mathrm{max}(|Q|)$. The dependence of $\mathrm{max}(|Q|)$ on $|R_1|/R_0$ and $z_0/L$ is plotted on Fig.~\ref{fig:sinusoidal_asymmetric_bifurcation} (d). While increasing $|R_1|/R_0$ generally increases the flow rate, there is an optimum value of $z_0/L$ beyond which increasing the shift $z_0$ between source and shape decreases the flow rate. Finally, we can see how the maximum values of $|Q|$ are maximized for different offsets $z_0/L$ depending on the value of $|R_1|/R_0$ in Fig.~\ref{fig:sinusoidal_asymmetric_bifurcation} (d).

For the symmetric ($z_0=0$) system, the pumping transition is captured entirely by the ratio $\tau_a/\tau_Q$. For the present asymmetric ($z_0 \neq 0$) case, while there is no pumping transition as before, there is still a transition between one and three real solutions of $Q$. This transition corresponds then to the point at which the pore may pump in both ways rather than just one, and is determined by Eq.~\eqref{eq:discrimant}. Plugging in the expressions for the timescales $\tau_a$ and $\tau_Q$ shows that they are not enough to capture this new transition. Indeed, one has to introduce a second advective timescale $\tilde{\tau}_Q$, 
\begin{equation}
    \tilde{\tau}_Q = \left( \frac{\pi^2}{L^2}  \frac{\mathcal{L}}{\beta \eta} \frac{R_1}{R_0} \frac{|\tilde{\xi}_1|}{\pi R_0^2} \right)^{-1/2},
\end{equation}
associated to the magnitude of the anti-symmetric part of the source $\tilde{\xi}_1$ rather than the symmetric part of the source $\xi_1$. With this new timescale, one identifies the condition for which the pore may pump in both ways as
\begin{equation}
 \left(    \frac{\tau_a}{\tau_Q} \right)^2 > 1 + \frac{3}{2^{2/3}} \left( \frac{\tau_a}{\tilde{\tau}_Q} \right)^{4/3}, \label{eq:pumpingTransition_asym}
\end{equation}
which is verified in Fig. \ref{fig:sinusoidal_asymmetric_adim}. To shed light on the meaning of Eq.~\eqref{eq:pumpingTransition_asym}, let us expand $\xi_1$ and $\tilde{\xi}_1$ to linear order in $z_0/L$. Expanding Eqs.~\eqref{eq:shiftedXi1} and~\eqref{eq:shiftedTildeXi1}, one obtains  
\begin{equation}
    \tau_{\tilde{Q}} \approx \tau_Q \left( \pi \frac{|z_0|}{L} \right)^{-1/2}
\end{equation}
and Eq.~\eqref{eq:pumpingTransition_asym} becomes
\begin{equation}
    \frac{4}{27 \pi^2} \left( \frac{\tau_a}{\tau_Q} \right)^2 \left[ 1 -  \left( \frac{\tau_a}{\tau_Q} \right)^{-2} \right]^3 >  \left( \frac{|z_0|}{L} \right)^{2}.
\end{equation}
In the strong pumping regime where $\tau_Q \gg \tau_a$, we obtain
\begin{equation}
    \sqrt{\frac{4}{27 \pi^2}}  \frac{\tau_a}{\tau_Q} >  \frac{|z_0|}{L}
\end{equation}
and
\begin{equation}
    \frac{\tau_a}{\tau_Q} \approx \frac{|Q|_{sym}}{R_0^2 L \tau_a^{-1}}, \label{eq:asym_cond1}
\end{equation}
where $Q_{sym}$ is the flow rate obtained when $z_0=0$ as per Eq.~\eqref{eq:GetQ4}. Thus, Eq.~\eqref{eq:asym_cond1} can be written as
\begin{equation}
  \sqrt{\frac{27}{4}}  |z_0| <  \frac{|Q|_{sym}}{\pi R_0^2} \tau_A, \label{eq:pumpingAnalysisShift}
\end{equation}
with the right-hand-side being roughly the distance a solute molecule is advected in a time $\tau_A$. Thus, pumping in both directions requires the flow rate to be high enough to carry solute from the maximum of the source function to the other side of the bottleneck (a distance $z_0$) before the solute is destroyed or diffuses away. In the case of a symmetric channel where $z_0=0$, Eq.~\eqref{eq:pumpingAnalysisShift} reduces to the condition $|Q_{sym}| > 0$, which is equivalent to Eq.~\eqref{eq:pumpingCond}.

\section{Fore-aft asymmetric shape}

\label{sec:NonSinusoidal}
\begin{figure*}[t]
	\centering
   \includegraphics[width=1.0\textwidth]{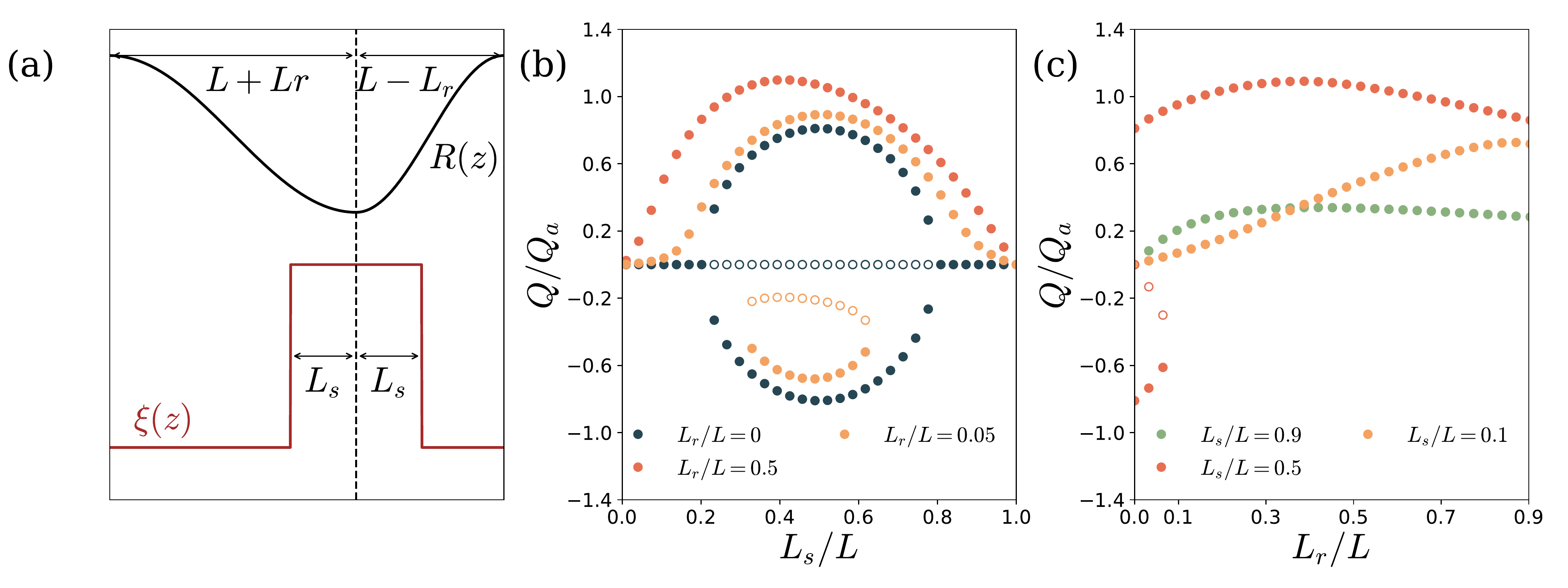}
   \caption{ (a) Diagram of the pore and source under study. (b) The possible steady state flow rates when varying $L_s/L$. Parameters are $(\mathcal{L}/\beta\eta) (\tau_{\chi}/L^5) = -10^{-6}$, $\xi_{max} L \tau_{\chi} = 10^5$, $R_{max}/L = 0.11$, $R_{min}/L = 0.09$, $D \tau_{\chi} /L^2 = 10^{-2}$. (c)  The possible steady state flow rates when varying $L_r/L$. Parameters same as panel (b). Open symbols correspond to unstable steady states and filled-in symbols correspond to stable steady states. \label{fig:ratchet_sym}} 
\end{figure*}

\begin{figure*}[t]
	\centering
   \includegraphics[width=1.0\textwidth]{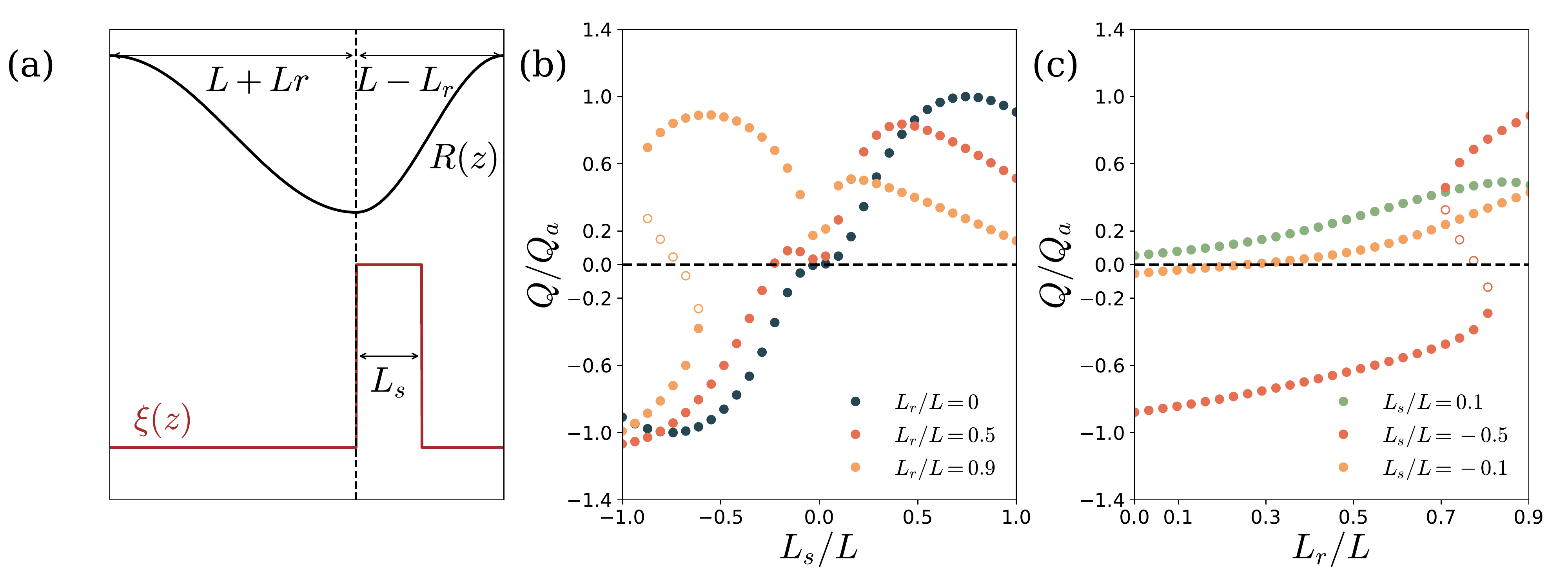}
      \caption{ (a) Diagram of the pore and source under study. (b) The possible steady state flow rates when varying $L_s/L$. Parameters are $(\mathcal{L}/\beta\eta) (\tau_{\chi}/L^5) = -10^{-6}$, $\xi_{max} L \tau_{\chi} = 10^5$, $R_{max}/L = 0.11$, $R_{min}/L = 0.09$, $D \tau_{\chi} /L^2 = 10^{-2}$. (c)  The possible steady state flow rates when varying $L_r/L$. Parameters same as panel (b). Open symbols correspond to unstable steady states and filled-in symbols correspond to stable steady states. \label{fig:ratchet_asym}} 
\end{figure*}

To further clarify the role of the pore shape on the pumping dynamics, we now study a pore with more complex geometry, described by a function $R(z)$ that is non-sinusoidal and asymmetric. We now define the radius as
\begin{equation}
    R(z)=
    \begin{cases}
      \frac{R_{max} + R_{min}}{2} - \frac{R_{max} - R_{min}}{2} \cos\left( \frac{\pi}{L+L_r} z\right), & \text{if}\ -L < z < L_r,  \\
      \frac{R_{max} + R_{min}}{2} - \frac{R_{max} - R_{min}}{2} \cos\left( \frac{\pi}{L-L_r} z\right), & \text{if}\  L_r \leq z < L ,
    \end{cases}
  \end{equation}
which has its minimum value at $z=L_r$, and exhibits a form similar to a smoothened sawtooth (see Fig.~\ref{fig:ratchet_sym} (a)). The asymmetry is controlled by the length $L_r$. Indeed, if $L_r = 0$, the function $R(z)$ becomes sinusoidal, and thus fore-aft symmetric. In order to assess the role of shape asymmetry, we need to take into account all the possible Fourier modes. Thus, we require a source function that contains all Fourier modes as well. To such effect, we introduce a step function source
\begin{equation}
\xi(z)=
    \begin{cases}
      \xi_{max} , & \text{if}\ z \in \Omega_{sym}(L_s,L_r),  \\
      0, & \text{otherwise,}
    \end{cases}
\end{equation} 
where $\Omega_{sym}$ is the region of points at a distance smaller than $L_s$ from the bottleneck, when accounting for periodic boundary conditions. A diagram of the source function can be found in Fig.~\ref{fig:ratchet_sym} (a). Both the source and the shape defined above exhibit non-zero Fourier components for all wavenumbers. This means the sum in Eq. ~\eqref{eq:centralQ} generally contains an infinite number of terms. We thus truncate the sum to include only the first fifteen terms, and we solve for $Q$ numerically. Including further terms does not significantly change the obtained solutions. We begin by investigating the effect of the extent of chemical activity in Fig.~\ref{fig:ratchet_sym} (b). If $L_s/L = 1$, the entire pore produces solute, and no pumping is obtained for any value of $L_r$. This is because a homogeneous source promotes a constant solute density, which results in zero slip velocity, and thus, no flow. This result holds up to linear order in $\delta R$, but may not hold to quadratic order or above. Indeed, Michelin et al. studied homogeneously-coated ratcheted channels, and found a flow rate that grows non-linearly with the corrugation height \cite{Michelin2015}.\\

In the symmetric case  ($L_r = 0$), pumping occurs only in a range of $L_s/L$ centered around $1/2$. Indeed, the flow rate exhibits two bifurcations and is symmetric with respect to $L_s/L=1/2$ (see Fig.~\ref{fig:ratchet_sym} (b)). When asymmetry in introduced ($L_r/L \neq 0 $), the symmetry with respect to $L_s/L=1/2$ is lost, and the two stable branches are no longer symmetric with respect to $Q =0$. The pitchfork bifurcations disappear and are replaced by imperfect bifurcations \cite{Strogatz2015}, as for the sinusoidal pore with $z_0 \neq 0$. Again there is a discontinuous jump with changing $L_s$ as the system jumps from the stable branch with $Q<0$ to the stable branch with $Q>0$. This discontinuity now arises from $R(z)$ being itself asymmetric, rather than from a shift between two symmetric functions $R(z)$ and $\xi(z)$, as was the case in the sinusoidal pore. For large enough asymmetries, however, there is only one solution, with the maximum flow rate depending on the values of $L_r/L$. This can be seen easily in Fig.~\ref{fig:ratchet_sym} (c), where the flow rate grows with $L_r/L$ until a maximum is reached, after which the flow rate decreases as $L_r/L \rightarrow 1$ (note that our assumption of weakly-varying $R(z)$ limits our approach to values of $L_r/L$ lower than about 0.9). Indeed, pumping is promoted when high gradients in solute density and in the radius coexist in the same position, as seen in Eq.~\eqref{eq:Qintegral}. If the drop in source strength (and thus, the drop in solute density) is located in a steep region, one observes higher values of $Q$, hence why an increasing value of $L_s/L$ leads to smaller values of the optimal value of $L_r/L$.

In the pores explored until now, the discontinuities in the flow rate manifested only for a prohibitively small degree of asymmetry. We now show that the discontinuities are a general phenomenon in asymmetric phoretic pores, and can be obtained for strongly asymmetric pores as well. To this effect, we consider solute production in only one of the halves of the pore as shown in Fig. ~\ref{fig:ratchet_asym} (a).
\begin{equation}
\xi(z)=
    \begin{cases}
      \xi_{max} , & \text{if}\ z \in \Omega_{asym}(L_s,L_r),  \\
      0, & \text{otherwise.}
    \end{cases}
\end{equation} 
Here, $\Omega_{asym}$ is the upper half of $\Omega_{sym}$ (accounting for periodic boundary conditions). Note that $L_s$ may now be negative (indicating that there is production only in a region in the weakly-sloped  half of the pore $z<L_r$ ). It can be seen in Fig.~\ref{fig:ratchet_asym} (b) that the symmetric pore ($L_r/L =0$) exhibits only one branch for $Q$, and it is antisymmetric. The maximum value of $|Q|$ is found at $L_s/L \approx \pm 0.8$. The antisymmetry is broken when pore asymmetry is introduced ($L_r/L > 0$). The maximum value of $Q$ is located at increasingly lower values of $L_s/L$ the more asymmetric the pore is (higher $L_r/L$).  All curves show no pumping ($Q=0$) when $L_s=0$. This is because the pore is chemically inert in this case. Remarkably, asymmetry induces a minimum at $Q=0$, with positive flow rates for both positive and negative $L_s$. This behavior acquires added complexity for sufficiently asymmetric pores ($L_r/L = 0.9$), where two stable branches can be found in a region of $L_s$. A second maximum arises for $L_s < 0$ and the flow rate there is even higher than the maximum for $L_s >0$. Therefore, sufficiently asymmetric pores pump optimally when chemical activity is present in the weakly-sloped half ($z < L_r$). This is in contrast with symmetric / weakly-asymmetric pores which pump optimally when chemical activity is present in the strongly-sloped half ($z >L_r$).
Figure~\ref{fig:ratchet_asym} (c) shows how for $L_s/L=-0.1$, it is still possible to obtain a stable non-pumping state ($Q=0$), even when the pore shape is asymmetric. Furthermore, it shows how, for the same value of $L_s/L$, the extent of the asymmetry dictates the direction of pumping. Two stable branches appear for $L_s/L=-0.5$, and a hysteresis loop is possible, now by varying the asymmetry $L_r/L$, rather than the active length $L_s/L$.

\section{Conclusions}

\label{sec:conclusions}

We have derived a model to describe self-diffusioosmotic transport of a chemically-reactive solution in an active pore. We have extended the classical self-diffusioosmotic/phoretic models to include the inverse chemical reaction, which consumes solute. By assuming a narrow, weakly-corrugated pore, we have reduced the 3D governing equations to an effective 1D system of advection-diffusion equations (Fick-Jacobs approximation \cite{Zwanzig1992, Reguera2001, Malgaretti2013}). The pore walls are coated in a catalytic material which triggers the production of solute. As the catalytic coating is spatially-varying, thus so is the composition of the
mixture. By tuning the extent of catalytic coating, as well as the pore's geometrical properties, one may harness diffusioosmosis to pump fluid across the pore. We have shown that this is true even in the case of a fore-aft symmetric pore, as a spontaneous symmetry breaking occurs when advection plays a large enough role. The model is analytically solvable for a sinusoidal pore, revealing that the appropriate control parameter is the ratio $\tau_a/\tau_Q$ of two timescales: the advective timescale $\tau_Q$, which determines how quickly a solute molecule is advected over the pore; and a second timescale $\tau_a$, which combines both the diffusive timescale, as well as the timescale associated to the consumption of solute. In the limit of negligible inverse chemical reaction, the control parameter $\tau_a/\tau_Q$ reduces to the P\'eclet number. In Appendix \ref{sec:hourglass}, we show that this control parameter can be used to approximately predict the onset of pumping for non-sinusoidal pores as well. 
Pores that break fore-aft symmetry, either due to their shape or due to their catalytic coating, pump as well, even if advection plays no role. Furthermore, asymmetric pores can still display bistability, where pumping in both directions is permitted. The bistability is such that the flow rate exhibits discontinuous jumps and hysteresis loops upon tuning the parameters that control the asymmetry. For sinusoidal pores asymmetrically coated in solute, it was found that the ratio $\tau_a/\tau_Q$ is not sufficient to describe this second transition between one and two possible directions. Instead, the appropriate control parameter requires the introduction of a fourth timescale, also associated to advective transport. 
Typically, experimental realizations of diffusioomosis make use of a hydrogen peroxide solution which is decomposed upon contact with platinum. We note that, while the inverse chemical reaction (formation of hydrogen peroxide from water and oxygen) is often not a decisive factor for experimental conditions, oxygen must nonetheless be removed from the pore to achieve a steady state. One possibility is to construct part of the pore's walls out of a porous gel, which allows oxygen to diffuse through \cite{Palacci2010}. In such a case, the sink constant should be understood as an effective parameter.
To obtain analytical results, we approximated the solute concentration as independent of the transverse direction. This approximation is better at smaller scales. Indeed, for an average radius of $1 \mu m$, the solute dynamics is expected to be reaction-limited \cite{Brown2014}, promoting a solute concentration that is homogeneous in the transverse direction. At larger scales, the Fick-Jacobs theory may still be used, given the appropriate choice of the source function, which should then be taken as an effective parameter. Such possibility arises from the flow field being entirely determined by the solute concentration near the wall. So long as that value is correct, it does not matter what values the solute distribution takes in the remaining pore volume.
Flows inside such pores show characteristic velocities of $v \leq 10 \mu m/s$ \cite{Ebbens2012}. We thus expect the flux associated with pumping active pores to be of magnitude of $10^{-2} pL / \mu m^2 / s$. Furthermore, symmetric active pores are expected to exhibit a pumping transition when the P\'eclet number $\approx 1$, and thus when $L \approx 10^2-10^3 \mu m$ \cite{Antunes2022}. We expect the discontinuous jumps in the flow rates of asymmetric pores to manifest at similar scales. An example of a catalytic material which exhibits pores of such sizes is that of metal foam catalysts \cite{He2012, Selvam2014}. We thus predict that advection-enabled instabilities will play a role in the functioning of such catalysts. To experimentally verify the discontinuous jumps and hysteresis in the flow rate, we thus propose experiments with platinum-coated pores containing hydrogen peroxide. These pores should showcase a length in the hundreds of micrometers. We suggest the fabrication of pores with slight asymmetry in the pore shape, as in Fig. \ref{fig:ratchet_sym} (a). To access both branches of Fig. \eqref{fig:ratchet_sym} (c), one may initially apply a pressure drop across the pore, to favor relaxation to either positive or negative values of $Q$ (depending on the sign of the pressure drop). Measuring the values of $Q$ for pores with increasing $L_r/L$, one may identify the end of a stable branch as the value of $L_r/L$ beyond which the pore only pumps in one direction, regardless of the pressure drop applied at the start of the experiment.

\vfill

\begin{acknowledgements}

We acknowledge funding by the Deutsche Forschungsgemeinschaft (DFG, German Research Foundation)—Project-ID 416229255—SFB 1411 and Project-ID 431791331—SFB 1452. Furthermore, we acknowledge the Helmholtz Association of German Research Centers (HGF) and the Federal Ministry of Education and Research (BMBF), Germany for supporting the Innovation Pool project “Solar H2: Highly Pure and Compressed”.

\end{acknowledgements}

\section*{Conflicts of Interest}
There are no conflicts of interest to declare.

\section*{Data Availability Statement}
The data that support the findings of this study is openly available in Zenodo at \url{http://doi.org/10.5281/zenodo.8233408}.

\section*{Author Contributions}
GCA: conceptualization, investigation, formal analysis, and writing - original draft. PM: conceptualization, formal analysis, supervision, and writing - review \& editing. JH: conceptualization, funding acquisition, supervision, writing - review \& editing.

\appendix

\section{The ratio $\tau_a/\tau_Q$ as a predictor of the pumping transition for a general symmetric pore}
\label{sec:hourglass}

\begin{figure*}[t]
	\centering
   \includegraphics[width=1.0\textwidth]{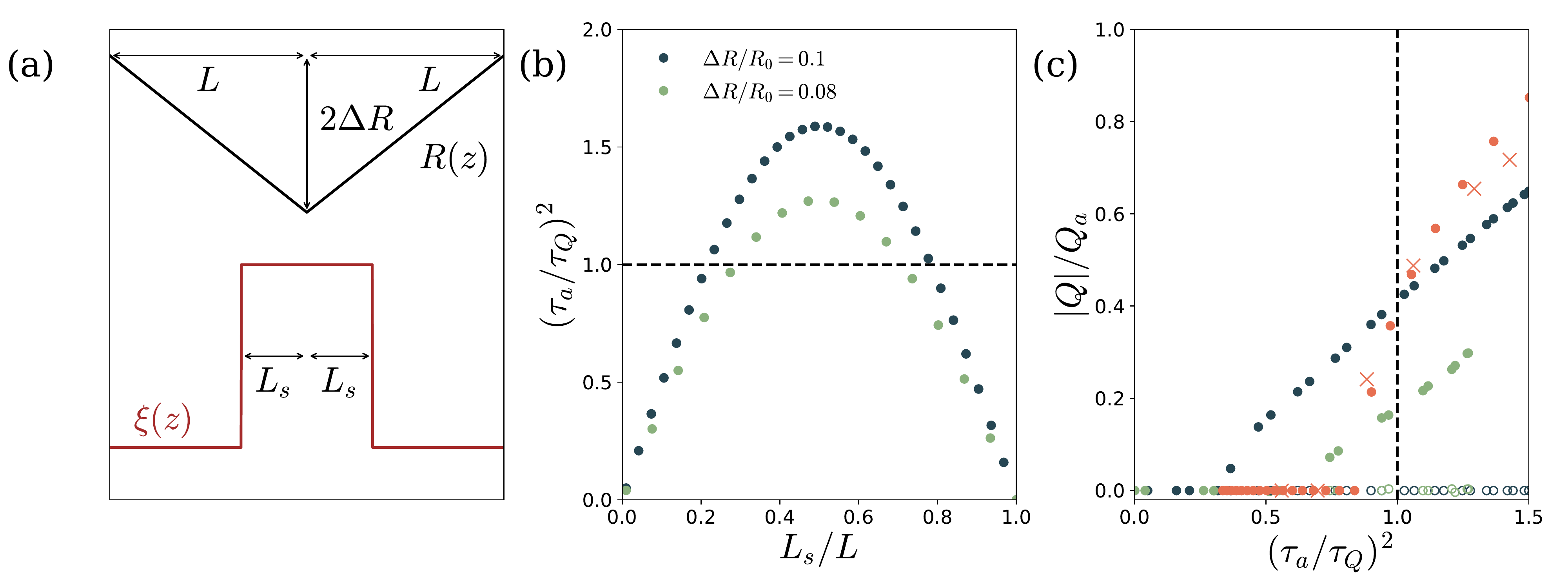}
   \caption{ (a) Diagram of the pore and source under study. (b) Variation of the adimensional number $\tau_a/\tau_Q$ with $L_s/L$. Parameters are $(\mathcal{L}/\beta\eta) (\tau_{\chi}/L^5) = -10^{-6}$, $\xi_{max} L \tau_{\chi} = 10^5$, $R_0/L = 0.1$, $D \tau_{\chi} /L^2 = 10^{-3}$. Fifteen Fourier modes were used for the computation. (c) Pumping rate as a function of the adimensional number $\tau_a/\tau_Q$. Orange crosses are simulation data from previous work obtained by varying the P\'eclet number\cite{Antunes2022}, and orange points are corresponding semi-analytical predictions upon fitting for $\xi_{max}$. Parameters are $(\mathcal{L}/\beta\eta) (\tau_{\chi}/L^5) = -1.3 \times 10^{-6}$, $\xi_{max} L \tau_{\chi} = 4.2 \times 10^7$, $R_0/L= 1.7$, $\Delta R/L= 0.29$. Twenty Fourier modes were used for the computation. \label{fig:hourglass}} 
\end{figure*}
We have understood the onset of pumping in symmetric sinusoidal pumps via two timescales ($\tau_a$ and $\tau_Q$). While $\tau_a$ does not depend on the height of the corrugation, the advective timescale $\tau_Q$ does. In principle, a generic pore is described by an infinite number of Fourier modes, and thus identifying the correct advective timescale needed to collapse the flow rate curves is not feasible. Nonetheless, often the first Fourier mode carries the largest weight, and so the question arises whether the ratio $\tau_a/\tau_Q$ may still provide an adequate approximation for the onset of pumping for a generic pore. For this test, we have chosen an hourglass-shaped pore with a step-function source (Fig.~\ref{fig:hourglass} (a)). As the width of the source function $L_s$ increases, $\xi_1$ initially grows and then decreases, leading to a non-monotonic curve for the dimensionless parameter $\tau_a/\tau_Q$ (Fig.~\ref{fig:hourglass} (b)). When plotting the normalized flow rate as a function of $\tau_a/\tau_Q$, the data does not collapse, as was expected. Nonetheless, the onset of pumping occurs for values of $\tau_a/\tau_Q$ close to unity, as seen in Fig.~\ref{fig:hourglass} (c). To further test the robustness of the current framework, we compare our theory to data from simulations \cite{Antunes2022} which solve the full hydrodynamics, as discussed in Appendix \ref{sec:sims}.  The theory shows a good agreement with the simulations, with the onset of pumping close to $\tau_a/\tau_Q = 1$. We thus conclude that $\tau_a/\tau_Q$ is a good estimator for the onset of pumping for a generic symmetric pore.

\section{Comparison with Lattice Boltzmann simulations}

\label{sec:sims}

\begin{figure}[t]
	\centering
   \includegraphics[width=.5\textwidth]{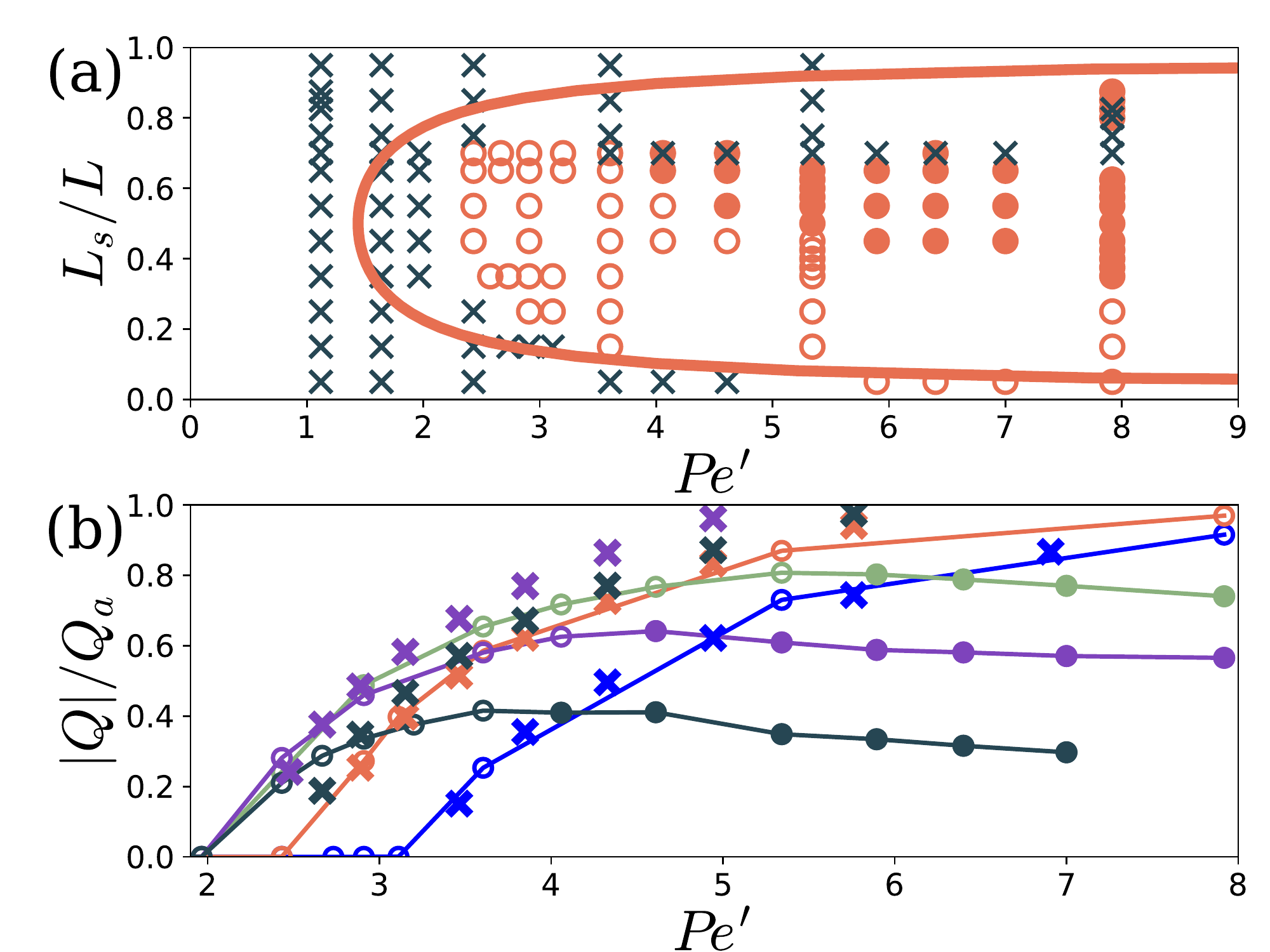}
   \caption{ (a) Classification of the asymptotic dynamics seen from the Lattice Boltzmann simulations of Ref. [\citenum{Antunes2022}] into non-pumping states ($\times$), steady pumping states ($\textcolor[HTML]{E76F51}{\circ}$), and oscillating states ($\textcolor[HTML]{E76F51}{\bullet}$). The line is the Fick-Jacobs theory's prediction for the onset of pumping for $\alpha_{fit}=17.1$. (b) Absolute value of state flow rate $|Q|$ as function of $Pe^{\prime}$ as obtained from the Fick-Jacobs theory for $\alpha_{fit}$ as per Table \ref{table:table} ($\times$) and Lattice Boltzmann simulations ($\circ, \bullet$). Lines are guides to the eye connecting simulation data. Colors indicate value of $L_s/L$: $0.15$ (\textcolor[HTML]{0000FF}{blue}) , $0.35$ (\textcolor[HTML]{E76F51}{orange}) , $0.45$ (\textcolor[HTML]{8AB17D}{green}), $0.55$ (\textcolor[HTML]{7E43BC}{purple}), $0.65$ (\textcolor[HTML]{264653}{gray}). Note that the Fick-Jacobs theory predicts the same value of $|Q|/Q_a$ for both $L_s/L = 0.45$ and $L_s/L = 0.55$.  Parameters are $(\mathcal{L}/\beta\eta) (\tau_{\chi}/L^5) = -1.3 \times 10^{-6}$, $\text{max}(\xi_{sims}) L^2 \tau_{\chi} = 3.8 \times 10^6$, $R_0/L= 1.7$, $\Delta R/L= 0.29$. Fifteen Fourier modes were used for the computation.  \label{fig:sims}}
\end{figure}

In order to judge the performance of the approximate theory derived in this work, we confront it with the corresponding numerical results of Reference[\citenum{Antunes2022}]. In these simulations, the pore shows an hourglass-shape and its walls are inhomogeneously coated with a catalytic material. As such, production of solute occurs on the pore walls, and only in a section of length $2L_s/L$ around the center of the channel (as in Fig. \ref{fig:hourglass} (a)). Figure \ref{fig:sims} (a) shows semi-quantitative agreement for the onset of pumping as a function of $L_s/L$ and the P\'eclet number $Pe^{\prime}$, defined as
\begin{equation}
    Pe^{\prime} = \frac{v^* L}{D},
\end{equation}
with $v^*$ being a typical velocity scale equal to $1.3 \nu \ L$ (with $\nu$ being the kinetic viscosity. Further details in Ref. [\citenum{Antunes2022}]). Since the simulations exhibit a pore that is wide ($R_0 /L = 1.7$), the simulations were performed out of the Fick-Jacobs theory's regime of validity. To accommodate for the resulting deviation from the simulations, we now treat the surface production rate $\xi_S$ as effective, and introduce a dimensionless fitting parameter $\alpha_{fit}$, such that
\begin{equation}
   \xi_S(z, L_s/L)= \alpha_{fit}(L_s/L) \xi_{sims}(z, L_s/L), \label{eq:xiSims}
\end{equation}
where $\xi_{sims}$ is the production rate per unit area used in the simulations. The value of $\alpha_{fit}$ is then extracted by fitting the theory curves $Q(Pe^{\prime}, L_s/L)$ to the curves obtained in the simulations.

\begin{table}[t]
	\centering
 \begin{tabular}{ |l |l |}
 \hline
     $L_s/L$  & $\alpha_{fit}$    \\
 \hline
 $0.15$ & $14.0 $ \\
 $0.25$ & $11.3 $ \\
 $0.45$ & $9.7 $ \\
 $0.55$ & $9.7$ \\
 $0.65$ & $9.6$ \\
  \hline
  \end{tabular}
 \caption{Values of $\alpha_{fit}$ for data in Fig. \ref{fig:sims} (b).}
 \label{table:table}
\end{table}

Comparing the flow rates obtained from the Fick-Jacobs theory and the Lattice Boltzmann simulations in Fig. \ref{fig:sims}, we see excellent fits for low values of $Pe^{\prime}$ and $L_s/L$. This fit is especially good in the regime in which the simulations exhibit a steady state, rather than sustained oscillations in the flow rate. As explained in Ref. [\citenum{Antunes2022}], said oscillations arise when the fluid relaxation time is finite. In the Fick-Jacobs theory, this is not the case, as we assume Stokes flow (Eq. \eqref{eq:app-stokes1}). Therefore, the oscillatory regime is beyond the scope of the current theory. Finally, it must be said that the active pore in the simulations is wide ($R_0/L = 1.7$), and thus out of the regime in which the approximations that lead to the Fick-Jacobs theory are valid. Nonetheless, we have shown that the theory may still show semi-quantitative agreement beyond this regime.

\section{Diagram summarizing model derivation}

\label{sec:diagram}

\begin{figure*}[t]
	\centering
		\begin{tikzpicture}[auto,node distance=1.5cm]
      \node[rectangle, minimum size = 11mm, draw, align = center, fill=white] (3DSE) {3D Stokes equation \\ Eq. \eqref{eq:app-stokes1} };
      \node[rectangle, minimum size = 11mm, draw, align = center, fill=white, right = 15 mm of 3DSE] (LUB) {$v_z$ as sum of \\ plug flow \\ and Poiseuille \\ flow \\ Eq. \eqref{eq:plugPlusPoi} };
      \path [-{Latex[length=2mm]}](3DSE) edge node [above] {$L \gg R_0$} node [below ]{$\nabla\cdot \bm{v} = 0$}  (LUB);
      \node[rectangle, minimum size = 11mm, draw, align = center, fill=white, right = 20 mm of LUB] (Q) {$Q$ as \\ function \\ of $v_0$ and $R$\\ Eq. \eqref{eq:Q} };
      \path [-{Latex[length=2mm]}](LUB) edge node [below]{ P.B.C. for $P$}   (Q);
      \node[rectangle, minimum size = 11mm, draw, align = center, fill=white, right = 25 mm of Q] (Q2) {$Q$ to \\ linear order \\ in $\delta R$ \\ Eqs. \eqref{eq:Qintegral} \\ and \eqref{eq:eqForP0}};
      \path [-{Latex[length=2mm]}](Q) edge node [above]{\begin{tabular} {l}                $R_0 \gg \delta R$, \\
                                                                                            $\partial_z R \ll 1$ 
                                                                                            \end{tabular} } node [below]  { \begin{tabular} {l}
                                                                                            P. B. C. for \\
                                                                                            $R$ and $\rho$ 
                                                                                            \end{tabular}}   (Q2);
      
            \node[rectangle, minimum size = 11mm, draw, align = center, fill=white, above = 35mm of 3DSE] (ADE) {3D advection-\\-diffusion \\equation \\ Eq. \eqref{eq:adv-diff} };
            \node[rectangle, minimum size = 11mm, draw, align = center, fill=white, right = 35mm of ADE] (INT) {Equation for \\ $\textbf{j}$ integrated \\ in the \\ cross-section \\ Eq. \eqref{eq:integratedJ} };
            \path [-{Latex[length=2mm]}](ADE) edge  node [above]  {$j_r(r > R)=0$} node [below] {\begin{tabular}{l} $ \mathbf{v} \rho$, $D\nabla \rho \gg$  \\
                                                                                         $ \gg \beta D\rho \nabla W$ 
                                    \end{tabular}}  (INT);
             
             \node[rectangle, minimum size = 11mm, draw, align = center, fill=white, above = 29mm of Q2] (P) {Effective 1D \\ equation for \\ integrated solute \\ concentration $p$ \\ Eq. \eqref{eq:toExpand_st} };
             \path [-{Latex[length=2mm]}](INT) edge node [above]{$\rho(r, z) \equiv \rho(z)$} node [below]  {\begin{tabular}{l}
                                                                                            $ L \gg R_0 \rightarrow$  \\
                                                                                            $ \rightarrow j_r \approx 0 ,$ \\
                                                                                            $Pe_r \ll 1,$ 
                                                                                            \end{tabular}}   (P);
          \path [-{Latex[length=2mm]}](P) edge node [sloped, above]{$R_0 \gg \delta R$} node [sloped, below]  {P. B. C. for $R$ and $\rho$}   (Q2);     
         \node[rectangle, minimum size = 11mm, draw, align = center, fill=white, right = 18mm of Q2] (FOU) { Polynomial \\ whose roots \\ are $Q$ \\ Eqs. \eqref{eq:centralQ}, \eqref{eq:FourierPStart}, \\ and \eqref{eq:FourierPEnd} };

         \path [-{Latex[length=2mm]}](Q2) edge node [above] { \begin{tabular}{l} P. B. C. \\ for $\hat{\xi}$ \end{tabular}} node [below] {\begin{tabular}{l} Fourier \\ transform \\ $\mathcal{P}$  and  $\hat{\xi}$ \end{tabular}}  (FOU);   
		\end{tikzpicture}
	   \caption{Diagram of the derivation of the model, including necessary assumptions. Arrows indicate which assumptions are required to proceed in the derivation. The acronym "P. B. C." stands for "periodic boundary conditions".}
    	\label{fig:diagram}
\end{figure*}
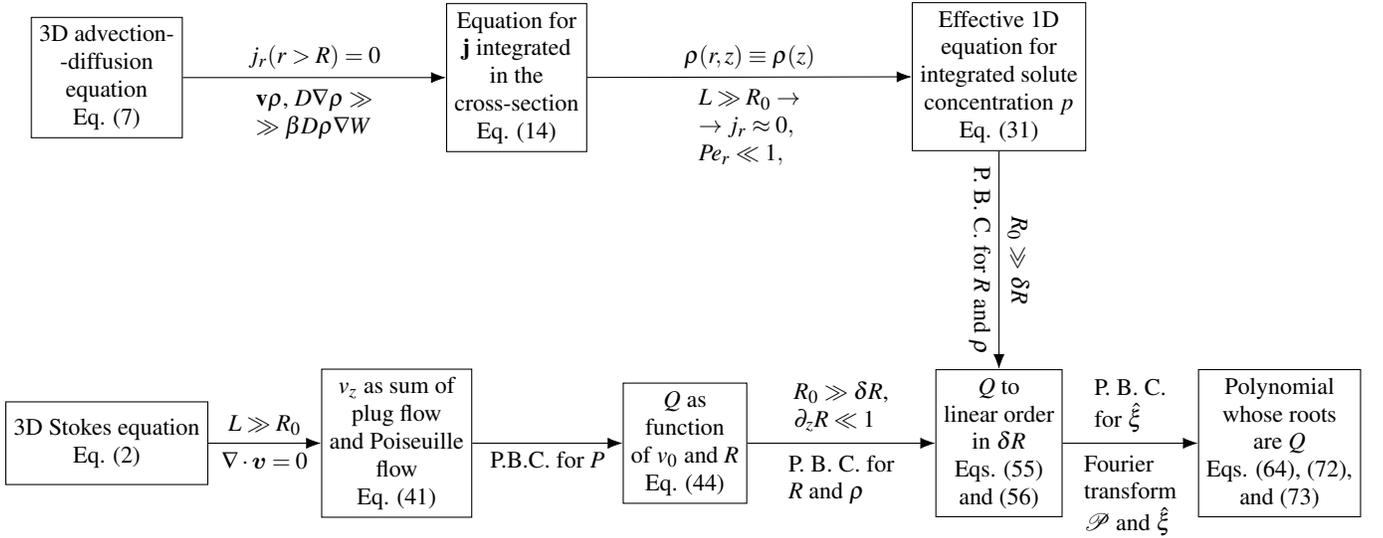

For the reader's convenience, we present a graphical summary of the model derivation in Fig. \ref{fig:diagram}.

\section{Linear Stability Analysis}
\label{sec:LSA}
We now examine the stability of the steady states described by Eqs.~\eqref{eq:centralQ},~\eqref{eq:FourierPStart}, and~\eqref{eq:FourierPEnd} by means of a linear stability analysis. We write
\begin{eqnarray}
\mathcal{P}(z,t) = \mathcal{P}_s(z) + \delta \mathcal{P}(z,t),\\
Q[\mathcal{P}(z,t)] = Q_s[\mathcal{P}_s(z)] + \delta Q[\delta \mathcal{P}(z,t)],
\end{eqnarray}
where $Q_s$ and $\mathcal{P}_s$ are the steady state flow rates and solute concentrations,  $\delta \mathcal{P} \ll \mathcal{P}_s$ is a perturbation to the steady state, and $\delta Q$ is the perturbation to the steady state flow rate. Therefore, Eqs.~\eqref{eq:eqForP0} and~\eqref{eq:Qintegral} become
\begin{align}
\dot{\mathcal{P}}_s(z) &=  D\partial_{z}^{2}\mathcal{P}_s(z) - \frac{Q_s}{\pi R_0^2}\partial_{z}\mathcal{P}_s(z) - \chi \mathcal{P}_s(z) + \xi(z),\\
Q_s &= \frac{\mathcal{L}}{\beta\eta } \frac{1}{R_0 L} \int\limits_{-L}^ L (\partial_z \mathcal{P}_s) \delta R(z) \  dz.
\end{align}
From Eq.~\eqref{eq:Qintegral}, we obtain
\begin{equation}
\delta Q(t) = \frac{\mathcal{L}}{\beta\eta } \frac{1}{R_0 L} \int\limits_{-L}^ L (\partial_z \delta \mathcal{P}(z,t)) \delta R(z) \  dx. \label{eq:Qintegral_LSA}
\end{equation}
We now expand Eq.~\eqref{eq:eqForP0} to linear order in $\delta \mathcal{P}(z)$, so that we get 
\begin{align}
 \delta \dot{\mathcal{P}}(z,t) = & D\partial_{z}^{2}\delta \mathcal{P}(z,t) - \frac{Q_s}{\pi R_0^2}\partial_{z}\delta \mathcal{P}(z,t) - \nonumber \\
 & - \frac{\delta Q(t)}{\pi R_0^2}\partial_{z} \mathcal{P}_s(z) - \chi \delta \mathcal{P}(z,t),\label{eq:mainP_LSA}
\end{align}
where we used $\mathcal{O}(\delta Q \partial_z \delta \mathcal{P}) = \mathcal{O}(\delta \mathcal{P}^2)$. We now decompose the perturbation $\delta \mathcal{P}(z,t)$ in Fourier modes in space,
\begin{align}
\delta \mathcal{P}(z,t) &= \sum\limits_{j=-\infty}^\infty q_j(t) e^{ik_jz}, \label{eq:defDeltaP_LSA}\\
q_{-j}(t) &= \bar{q}_j(t), \label{eq:defDeltaP3_LSA}
\end{align}
where $k_j = (\pi /L) j$. The time dependence of $\delta \mathcal{P}(z,t)$ is encoded in the set of amplitudes $\{q_j(t)\}$, which are complex and wavenumber-dependent. Equation~\eqref{eq:defDeltaP3_LSA} ensures that $\delta \mathcal{P}(z,t)$ is a real function. We further decompose $\mathcal{P}_s(z)$ in its Fourier components,
\begin{align}
\mathcal{P}_s(z) &= \sum\limits_{j=-\infty}^\infty B_j e^{ik_jz} , \label{eq:defP0_LSA}\\
B_{-j} &= \bar{B}_j, \label{eq:defP0_2_LSA}\\
B_{j} &= \frac{\mathcal{P}_j}{2}-i\frac{\tilde{\mathcal{P}}_j}{2} \text{ , } j>0, \label{eq:defP0_3_LSA}\\
B_0 &= \mathcal{P}_0
\end{align}
where $\mathcal{P}_0$, $\mathcal{P}_j$ and $\tilde{\mathcal{P}}_j$ are the coefficients in Eqs.~\eqref{eq:FourierPStart} and~\eqref{eq:FourierPEnd}. We can now write 
\begin{align}
&\frac{\delta Q(t)}{\pi R_0^2} \partial_z \mathcal{P}_s(z) = \nonumber \\
&= \frac{1}{\pi R_0^2} \frac{\mathcal{L}}{\beta \eta} \sum\limits_{j=-\infty}^\infty  \sum\limits_{j\prime=-\infty}^\infty k_{j\prime} q_j(t) B_{j\prime} e^{i k_{j \prime}z} [\Gamma_j + i\Theta_j].
\end{align}
Plugging the above equation into Eq.~\eqref{eq:mainP_LSA}, we obtain
\begin{align}
\dot{q}_j(t) &= \sum\limits_{j=-\infty}^\infty M_{ j  j\prime} q_{j\prime}(t), \label{eq:dynSysLSA}\\
M_{jj\prime} &= -\left[ Dk_j^2 + i\frac{Q_s}{\pi R_0^2}k_j + \chi  \right]\delta_{j j\prime} - \frac{\mathcal{L}}{\beta \eta} \frac{k_j B_j}{\pi R_0^2} (\Gamma_{j\prime} + i\Theta_{j\prime}). \label{eq:LSA_M}
\end{align}
Note that the matrix $M$ is not diagonal in general. In simple cases, it is possible to obtain the eigenvalues of $M$ analytically, and thus to determine the stability of each steady state. In the most general case, we determine stability by solving the dynamical system of equation~\eqref{eq:dynSysLSA} using forward Euler time integration. Starting from a random set of amplitudes at time $t=0$, we determine stability by whether the point $\{q_j\}(t)$ diverges from the origin given a certain time.

Consider the case where $B_j$ is zero or both $\Gamma_j$ and $\Theta_j$ are zero for a specific value of $j$. This ensures that the off-diagonal element
\begin{equation}
 M_{j \neq j\prime} = -\frac{\mathcal{L}}{\beta \eta} \frac{k_j B_j}{\pi R_0^2} (\Gamma_{j\prime} + i\Theta_{j\prime}),
\end{equation}
is zero. This is the case if the source function $\xi(x)$ and the shape $R(x)$ are both of the form $a_1 + a_2\cos(k_r x) +a_3 \sin(k_r x)$. Then $B_j, \Gamma_j, \Theta_j =0$, for all values of $j \neq r$. Under these conditions, the wavenumbers are decoupled, and we obtain a matrix equation for $q_j$:
\begin{gather}
 \begin{bmatrix} \dot{q}_j  \\ \dot{q}_{-j} \end{bmatrix}
 =
  \begin{bmatrix}
  M^{\prime}_{11}  &
  0 \\
  0 &
  M^{\prime}_{22} 
   \end{bmatrix}
    \begin{bmatrix} q_j  \\ q_{-j} \end{bmatrix},
\end{gather}
where
\begin{align}
M^{\prime}_{11} = M_{jj} &= -\left[Dk_j^2 + i\frac{Q_s}{\pi R_0^2}k_j + \chi\right] , \label{eq:Mprime11}\\
M^{\prime}_{22} = M_{-j-j} &= -\left[Dk_j^2 - i\frac{Q_s}{\pi R_0^2}k_j + \chi\right].\label{eq:Mprime22}
\end{align}
The real part of $M^{\prime}_{11}$ and $M^{\prime}_{22}$ is negative and therefore these modes are stable. The only mode that can possibly be unstable is that corresponding to $k_r$, the one that describes the source and shape. We obtain for $j=r$,
\begin{gather}
 \begin{bmatrix} \dot{q}_r  \\ \dot{q}_{-r} \end{bmatrix}
 =
  \begin{bmatrix}
  M_{11}  &
  M_{12} \\
  M_{21} &
  M_{22} 
   \end{bmatrix}
    \begin{bmatrix} q_r  \\ q_{-r} \end{bmatrix},
\end{gather}
where
\begin{align}
M_{11} = M_{rr} &= -\left[Dk_r^2 + i\frac{Q_s}{\pi R_0^2}k_r + \chi\right] - \nonumber \\
                &- \frac{\mathcal{L}}{\beta \eta} \frac{k_r B_r}{\pi R_0^2} (\Gamma_r + i\Theta_r), \label{eq:M11}\\
M_{12} = M_{r-r} &=  -  \frac{\mathcal{L}}{\beta \eta} \frac{k_r B_r}{\pi R_0^2} (-\Gamma_r + i\Theta_r),\\
M_{21} = M_{-r-r} &= -  \left(-\frac{\mathcal{L}} {\beta \eta} \right) \frac{k_r \bar{B}_r}{\pi R_0^2} (\Gamma_r + i\Theta_r) = \bar{M}_{12},\\
M_{22} = M_{-r-r} &= -\left[Dk_r^2 - i\frac{Q_s}{\pi R_0^2}k_r + \chi\right] - \nonumber \\
                  &- \left(- \frac{\mathcal{L}}{\beta \eta} \right) \frac{k_r \bar{B}_r}{\pi R_0^2} (-\Gamma_r + i\Theta_r) = \bar{M}_{11}.\label{eq:M22}
\end{align}
We now calculate the eigenvalues $\lambda$ from
\begin{equation}
(M_{11} - \lambda)(M_{22}-\lambda) - M_{12}M_{21} = 0,
\end{equation}
which results in
\begin{equation}
2\lambda = M_{11} + M_{22} \pm \sqrt{(M_{11}+M_{22})^2 -4(M_{11}M_{22}-M_{12}M_{21})}.
\end{equation}
The first term is calculated as
\begin{equation}
M_{11}+M_{22} = 2Re(M_{11}) = -2[Dk_r^2 + \chi] - \frac{\mathcal{L}}{\beta \eta} \frac{k_r}{\pi R_0^2} (\mathcal{P}_r \Gamma_r + \tilde{\mathcal{P}}_r \Theta_r).
\end{equation}
Furthermore,
\begin{align}
&M_{11}M_{22}-M_{12}M_{21} = M_{11}\bar{M}_{11} - M_{12}\bar{M}_{12} = \nonumber \\
& =Re^2(M_{11}) + Im^2(M_{11}) -Re^2(M_{12}) - Im^2(M_{12}),
\end{align}
and from Eq.~\eqref{eq:defP0_3_LSA} combined with Eqs.~\eqref{eq:M11}-\eqref{eq:M22}, we obtain
\begin{align}
Re(M_{11}) &= -[Dk_r^2 + \chi] - \frac{\mathcal{L}}{\beta \eta} \frac{k_r}{2\pi R_0^2} (\mathcal{P}_r \Gamma_r + \tilde{\mathcal{P}}_r \Theta_r),\\
Im(M_{11}) &= -\frac{Q_s}{\pi R_0^2}k_r - \frac{\mathcal{L}}{\beta \eta} \frac{k_r}{2\pi R_0^2} (\mathcal{P}_r \Theta_r - \tilde{\mathcal{P}}_r \Gamma_r),\\
Re(M_{12}) &=  - \frac{\mathcal{L}}{\beta \eta} \frac{k_r}{2\pi R_0^2} (-\mathcal{P}_r \Gamma_r + \tilde{\mathcal{P}}_r \Theta_r),\\
Im(M_{12}) &=  - \frac{\mathcal{L}}{\beta \eta} \frac{k_r}{2\pi R_0^2} (\mathcal{P}_r \Theta_r + \tilde{\mathcal{P}}_r \Gamma_r).
\end{align}
Finally,
\begin{align}
&M_{11}M_{22}-M_{12}M_{21} =  [Dk_r+\chi]^2 + \left[ \frac{Q_s}{\pi R_0^2}k_r  \right]^2 - \nonumber \\
-& \left( \frac{\mathcal{L}}{\beta \eta} \right) \frac{k_r}{\pi R_0^2} \left\{ \mathcal{P}_r \left[(Dk_r^2 + \chi) \Gamma_j + \frac{Q_s}{\pi R_0^2}k_r \Theta_j\right] + \nonumber \right.\\
&+ \left. \tilde{\mathcal{P}}_r \left[ (Dk_r^2 + \chi) \Theta_r -  \frac{Q_s}{\pi R_0^2} k_r \Gamma_r \right] \right\},   
\end{align}
where the identity $(a+b)^2 - (a-b)^2 = 4ab$ is useful. Then, 
\begin{align}
&(M_{11}+M_{22})^2 -4(M_{11}M_{22}-M_{12}M_{21}) = \nonumber \\
&\left[\frac{\mathcal{L}}{\beta \eta} \frac{k_r}{\pi R_0^2}(\mathcal{P}_r \Gamma_r + \tilde{\mathcal{P}}_r \Theta_r) \right]^2 -8\left[ \frac{Q_s}{\pi R_0^2} k_r \right]^2,
\end{align}
where we have used Eq.~\eqref{eq:centralQ}. The eigenvalues can now be written  as 
\begin{align}
2 \lambda = &-2[Dk_r^2 + \chi] - \frac{\mathcal{L}}{\beta \eta} \frac{k_r}{\pi R_0^2} (\mathcal{P}_r \Gamma_r + \tilde{\mathcal{P}}_r \Theta_r) \pm \nonumber \\
&\pm \sqrt{\left[\frac{\mathcal{L}}{\beta \eta} \frac{k_r}{\pi R_0^2}(\mathcal{P}_r \Gamma_r + \tilde{\mathcal{P}}_r \Theta_r) \right]^2 -8\left[ \frac{Q_s}{\pi R_0^2} k_r \right]^2},
\end{align}
with the condition for stability $Re(\lambda)<0$ returning
\begin{eqnarray}
 2[Dk_r^2 + \chi] > - \frac{\mathcal{L}}{\beta \eta} \frac{k_r}{\pi R_0^2} (\mathcal{P}_r \Gamma_r + \tilde{\mathcal{P}}_r \Theta_r), \\
\left[\frac{\mathcal{L}}{\beta \eta} \frac{k_r}{\pi R_0^2}(\mathcal{P}_r \Gamma_r + \tilde{\mathcal{P}}_r \Theta_r) \right]^2  < 8\left[ \frac{Q_s}{\pi R_0^2} k_r \right]^2,
\end{eqnarray}
for relaxation to steady state with ringing around the steady state value ($Im(\lambda) \neq 0$), and 
\begin{eqnarray}
 2[Dk_r^2 + \chi] > - \frac{\mathcal{L}}{\beta \eta} \frac{k_r}{\pi R_0^2} (\mathcal{P}_r \Gamma_r + \tilde{\mathcal{P}}_r \Theta_r) + \nonumber \\
 +\sqrt{\left[\frac{\mathcal{L}}{\beta \eta} \frac{k_r}{\pi R_0^2}(\mathcal{P}_r \Gamma_r + \tilde{\mathcal{P}}_r \Theta_r) \right]^2 -8\left[ \frac{Q_s}{\pi R_0^2} k_r \right]^2}, \\
\left[\frac{\mathcal{L}}{\beta \eta} \frac{k_r}{\pi R_0^2}(\mathcal{P}_r \Gamma_r + \tilde{\mathcal{P}}_r \Theta_r) \right]^2  > 8\left[ \frac{Q_s}{\pi R_0^2} k_r \right]^2,
\end{eqnarray}
for relaxation to steady state with no ringing ($Im(\lambda) = 0$).

\bibliography{refs,ref_paolo}

\end{document}